\documentclass[a4paper,12pt]{article}
\usepackage[dvips]{graphics}
\usepackage{epsfig}
\usepackage{psfrag}
\usepackage[psamsfonts]{amssymb}
\usepackage{amsmath}
\usepackage{indentfirst}
\usepackage{amssymb}
\usepackage{wrapfig}
\usepackage{natbib}
\usepackage{url}
\usepackage[usenames,dvipsnames]{color}

\newcommand{\E}{\mathrm{E}}
\newcommand{\Var}{\mathrm{Var}}

\begin{document}

\begin{center}
\Large{Super-exponential endogenous bubbles in an equilibrium model of rational and noise traders}
\end{center}

\begin{center}
\small{Taisei Kaizoji$^1$, Matthias Leiss$^2$, Alexander Saichev$^{2,3}\dagger$  and Didier Sornette$^{2,4}$\\
\vskip 0.3cm
$^1$ Graduate School of Arts and Sciences, International Christian University\\
3-10-2 Osawa, Mitaka, Tokyo 181-8585, Japan\\
$^2$ ETH Zurich, Department of Management, Technology and Economics\\
Kreuzplatz 5, CH-8032 Zurich, Switzerland\\
$^3$ Nizhni Novgorod State University,
Department of Mathematics, Russia\\
$^4$ Swiss Finance Institute,
c/o University of Geneva\\
40 blvd. Du Pont dÕArve, CH 1211 Geneva 4, Switzerland\\
\vskip 0.3cm
kaizoji@icu.ac.jp,  mleiss@ethz.ch, dsornette@ethz.ch\\
$\dagger$ deceased 8 June 2013
\vskip 0.5cm

}
\end{center}

\vskip 0.5cm

\noindent {\bf Abstract}: We introduce a model of super-exponential financial bubbles 
with two assets (risky and risk-free), in which rational investors and noise 
traders co-exist. Rational investors
form expectations on the return and risk of a risky asset and 
maximize their constant relative risk aversion expected utility with respect to their allocation
on the risky asset versus the risk-free asset. Noise traders are subjected to social
imitation and follow momentum trading. 
Allowing for random time-varying
herding propensity, we are able to reproduce several well-known stylized facts
of financial markets such as a fat-tail distribution of returns and volatility clustering.
In particular, we observe transient faster-than-exponential bubble growth with 
approximate log-periodic behavior and give analytical arguments why this follows 
from our framework.
The model accounts well for the behavior of traders and for the price
dynamics that developed during the dotcom bubble in 1995-2000.
Momentum strategies are shown to be transiently profitable, supporting
these strategies as enhancing herding behavior.

\vskip 1cm

\noindent
{\bf keywords}: noise traders, financial bubbles, faster-than-exponential growth, social imitation, momentum trading,
dotcom bubble

G01 -  Financial crises\\
G17 - Financial Forecasting

\pagebreak

\section{Introduction}

The very existence of financial bubbles has been a controversial and elusive subject.
Some have argued that financial bubbles play a huge role in the global economy,
affecting hundreds of millions of people (Kindleberger, 1978; Shiller, 2000; Sornette , 2003). 
Others have basically ignored or refuted their possibility (Fama, 1998). Moreover, 
until recently, the existence of such bubbles, much less
their effects, have been ignored at the policy level. Finally, only after the most recent historical global
financial crisis, officials at the highest level of government and academic
finance have acknowledged the existence and importance of identifying and understanding bubbles.
the President of the Federal Reserve Bank of New York, William C. Dudley, stated in April 2010
``what I am proposing is that we try|try to identify bubbles in real time, try to develop
tools to address those bubbles, try to use those tools when appropriate to limit the size of
those bubbles and, therefore, try to limit the damage when those bubbles burst.''
Such a statement from the New York Fed representing, essentially, the monetary policy of the United
States governmental banking system would have been, and, in some circles, still is, unheard of. This,
in short, is a bombshell and a wake-up call to academics and practitioners. Dudley exhorts to try to
develop tools to address bubbles.

But before acting against bubbles, before even making progress in 
ex-ante diagnosing bubbles, one needs to define what is a bubble.
The problem is that the ``econometric detection of asset price bubbles cannot be achieved with
a satisfactory degree of certainty. For each paper that finds evidence of bubbles, there
is another one that fits the data equally well without allowing for a bubble. We are
still unable to distinguish bubbles from time-varying or regime-switching
fundamentals, while many small sample econometrics problems of bubble tests
remain unresolved.'' summarizes Gurkaynak (2008) in his review paper. 

Let us start with the rather generally accepted stylized fact that,
in a period where a bubble is present, the stock return exhibits 
transient excess return above the long-term historical average, 
giving rise to what could be termed a ``bubble risk premium puzzle''.
For instance, as we report in the empirical section, the valuation of 
the Internet stock index went from a reference value $100$ in January 1998
to a peak of $1400.06$ in March 9, 2000, corresponding to an annualized
return of more than $350\%$ ! A year and a half later, the Internet stock 
valuation was back at its pre-1998 level.  Another stylized
fact well represented during the dotcom bubble is the highly intermittent
or punctuated growth of the stock prices, with super-exponential
accelerations followed by transient corrections, themselves followed
by further vigorous rebounds (Johansen and Sornette, 2010; Sornette and Woodard, 2010).
Bubbles are usually followed by crashes, in an often tautological 
logic resulting from the fact that the existence of a crash is usually taken as the ex-post signature
of the bubble, as summarized by A. Greenspan (2002):
``We, at the Federal Reserve... recognized that, despite our suspicions, it was very
difficult to definitively identify a bubble until after the fact, that is, when its
bursting confirmed its existence...'' More optimistically but
still controversial, recent systematic econometric studies
have shown that it is possible to relate objectively an anomalous 
transient excess return and the subsequent crash 
(Sornette, 2003; Johansen and Sornette, 2010; Sornette et al., 2011).

The recent finance literature has evolved to increasingly recognize the evidence of bubbles which is defined as deviations from fundamental value. One important class of theories is related to {\it noise traders} (also referred to as positive-feedback investors). The term ``noise traders'' was introduced first by Kyle (1985) and Black (1986) to describe irrational investors.  Thereafter, many scholars exploited this concept to extend the standard models by introducing the simplest  possible heterogeneity in terms of two interacting populations of rational and irrational agents. One can say that the one-representative-agent theory is being progressively replaced by a two-representative-agents  theory, analogously to the progress from the one-body to the two-body problems in physics. It has been often explained that markets bubble and crash in the absence of significant shifts in economic fundamentals is often explained to occur when herders such as noise traders deliberately act against their private information and follow the crowd. \par

De Long, Shleifer, Summers and Waldmann (1990a, 1990b) proposed the first model of market bubbles and crashes which  exploits this idea of the possible role of noise traders following positive feedback strategies or 
momentum investment strategies in the development of bubbles. They showed a possible mechanism for why asset prices may deviate from the fundamentals over long time periods. 
The key point is that trading  between rational arbitrageurs and noise traders gives rise to bubble-like price patterns. In their model, rational speculators destabilize prices because their trading triggers positive feedback trading by noise traders. Positive feedback trading by noise traders leads to a positive  auto-correlation of returns at short horizons. Eventually, arbitrage by rational speculators will pull the prices back to fundamentals. 
Their arbitrage trading leads to a negative autocorrelation of returns at longer horizons.  

Their work was followed by a number of empirical studies on positive feedback tradings. 
The influential empirical evidence on positive feedback trading came from the works of De Bondt and Thaler (1985), and Jegadeesh and Titman (1993, 2001), which established that stock returns exhibit momentum behavior at intermediate horizons, and reversals at long horizons. That is, strategies which buy stocks that have performed well in the past and sell stocks that have performed poorly in the past generate significant positive returns over 3- to 12- month holding periods. However, stocks that perform poorly in the past perform better over the next 3 to 5 years than stocks that perform well in the past. Behavioral models that explain the coexistence of intermediate horizon momentum and long horizon reversals in stock returns are proposed by Barberis, Shleifer, and Vishny (1998), Daniel, Hirshleifer, and Subrahmanyam (1998),  and Hong and Stein (1999). \par

The behavior of investors who are driven by group psychology, so-called {\it interacting agents}, and the aggregate behavioral outcomes, have also been studied using frameworks suggested by Weidlich and Haag (1983), Blume (1993; 1995), Brock (1993), Durlauf (1997; 1999), Kirman (1993), Brock and Durlauf (2000),  Aoki and Yoshikawa (2007), Chiarella, Dieci and He (2009) and Hommes and Wagener (2009).
Phan et al. (2004) summarize the formalism starting with different implementation of the agents' decision processes whose aggregation is inspired from statistical mechanics to account for social influence in individual decisions. Lux (1995), Lux and Marchesi (1999), Brock and Hommes (1999), Kaizoji (2000, 2010), and Kirman and Teyssiere (2002) have developed related models in which agents' successful forecasts reinforce the forecasts. Such models have been found to generate swings in opinions, regime changes and long memory. An essential feature of these models is that agents are wrong for a fraction of the time but, whenever they are in the majority, they are essentially right by a kind of self-fulfilling prophecy. Thus, they are not systematically irrational (Kirman, 1997). Sornette and Zhou (2006) showed how irrational Bayesian learning added 
to the Ising model framework reproduces the stylized facts of financial markets. Harras and Sornette (2011) showed how over-learning from lucky runs of random news in the presence of social imitation may lead to endogenous bubbles and crashes. \par

Here, we follow this modelling path and develop a model of the pricing mechanism 
and resulting dynamics of two co-existing classes of assets, a risky asset 
representing  for instance the Internet sector during the dotcom bubble and 
a risk-free asset, in the presence of two types of investors having different 
opinions concerning the risky asset (Harrison and Kreps, 1978; Scheinkman 
and Xiong, 2003). The first type of traders is a group 
of rational investors who maximize their expected utility. The second 
type of traders is a group of ``noise traders'' who trade only the risky 
asset by using heuristics such as past momentum and social imitation. 
The noise traders do not consider the fundamentals, while the rational 
investors allocate their wealth based on their expectation of the future 
returns and risks of the risky asset.  \par

Our framework combines elements from various 
groundbreaking works. The setup of noise traders follows closely 
Lux and Marchesi (1999), where an opinion index determined by 
past momentum and social imitation describes the prevailing investment
behavior among this group. The description of rational investors is related
to Brock and Hommes (1999) and to Chiarella, Dieci and He (2009). In particular,
we employ a utility function with constant relative risk aversion, as this is a realistic 
choice in a growing economy. \par

One important ingredient that we introduce here is that we do not allow our agents
to switch their investment behavior from rational to noise trading or vice versa. This reflects
the empirical fact that many large institutional investors such as pension funds 
have to follow strict guidelines on how to split their portfolio on assets of different 
risk classes. In previous models, the occurrence of a bubble was related to a 
convergence of a large fraction of traders on noise trading, see for example 
Lux and Marchesi (1999). Instead of strategy switching, we account for the
volatility of the imitation propensity of noise traders by assuming that it 
fluctuates randomly around some anchoring value as in (Stauffer and Sornette, 1999; Harras et al., 2012).
By keeping track of the agents' wealth levels, we are able 
to explain bubbles only with the transient increasing influence of noise traders on the market
price during an appreciation of the risky asset. While its price is rising, noise 
traders believing in momentum tend to invest more in the risky asset and thus become
richer, thereby gaining more importance. The noise traders' belief is further
reinforced by social imitation, which becomes self-fulfilling.
This, in turn, has destabilizing effects leading to 
an increase in the volatility and usually finishes in a crash when the prevailing 
opinion switches to pessimistic. \par

Our simple setup without strategy switching
reproduces several stylized facts of financial markets. The distribution
of returns is fat-tailed with a realistic power law exponent. Furthermore, signed returns are 
characterized by a fast-decaying autocorrelation, while the autocorrelation function for
absolute returns has a long memory (volatility clustering). 
We show theoretically and by simulations that bubbles start with a phase
of transient faster-than-exponential growth. Documented extensively for
bubbles in real markets (see for example Sornette et al. 2009, Jiang et al. 2010 and Yan et al. 2012)
and recently observed in lab experiments (H\"usler, Sornette and Hommes 2013), 
so far this behavior has been rarely discussed in the context of agent-based models. 
A first instantiation is found in (Corcos et al., 2002), in a much simplified model
of imitative and contrarian agents. The present model is one of the first in which 
we can provide a transparent analytical explanation for the 
existence of a transient faster-than-exponential growth.
Moreover, we observe approximate log-periodic behavior during the rise of a bubble, 
that can result from the nature of the fluctuations of the opinion index. 
While many of the ingredients and conditions used in our agent-based model may be found
in various forms  in some previous agent-based models, none have documented
explicitly the important transient super-exponential behavior associated with bubbles, nor
explained qualitatively or quantitatively the underlying mechanisms and the 
coexisting salient stylized facts.

The paper is organized as follows: the basic model is 
presented in Section 2 and Section 3 and analyzed theoretically in Section 4. 
Numerical simulations of the model are performed and the results are discussed in Section 5,
together with a discussion and quantitative characterization of the price
dynamics, its returns and momentum strategies during the dotcom bubble
from 1998 to 2000. We conclude in Section 6.

\section{Set-up of the model of an economy made of rational and of noise investors}

We consider fixed numbers $N_{\rm rational}$ of rational arbitrageurs and $N_{\rm noise}$
of noise investors who trade the same risky asset, represented here for simplicity by a single
representative risky asset fund. 
The arbitrageurs diversify between the risky asset and a risk-free asset on the basis 
of the maximization of their constant relative risk aversion expected utility, 
based  on their expectation of the returns and variance of the risky asset over the next period.
The noise traders use technical and social indicators, such as price momentum
and social imitation to allocate their wealth. A dynamically evolving fraction of them buys the risky asset 
while others stay out of the risky asset and have their wealth invested in the risk-free asset. 

In the next subsection \ref{trjukj}, we solve the standard allocation problem for the 
rational investors that determines their demand for the risky asset.
Then, in subsection \ref{hj6j8j}, the general ingredients controlling the dynamics of the demand of noise traders
are developed.

\subsection{Allocation equation for the rational investors \label{trjukj}}

The objective of the $N_{\rm rational}$ rational investors is assumed
to be the maximization at each time $t$ of the expected utility of 
their expected wealth $W_{t+1}$ at the next period, thus following  
Chiarella et al. (2009) and Hommes and Wagener (2009).
To perform this optimization,
they select at each time $t$ a portfolio mix of the risky asset and of the risk-free asset that they 
hold over the period from $t$ to $t+1$. Such one-period ahead optimization
strategy can be reconciled with underlying expected utility maximizing 
stories as given for example in (Brock and Hommes 1997, 1998; 
Chiarella et al., 2009; Boswijk, Hommes and Manzan 2007, Hommes and Wagener, 2009).

The rational investors are assumed to be identical, so that we can consider the behavior of one representative 
rational investor hereafter. We shall assume that rational investors are myopic mean-variance maximizers,
which means that only the expected portfolio value and its variance impact their allocation. 
We denote $P_t$ the price of the risky asset and $X_t$ the
number of risky assets that the representative rational investor
holds at instant $t$. We also assume that the risky asset
pays a dividend $d_t$ at each period $t$. 
Similarly, $P_{ft}$ and $X_{ft}$ correspond to the 
price and number of a risk-free asset held by the rational agent.
The risk-free asset is in perfectly elastic supply and pays 
a constant return $R_f$. Thus, at time $t$, the wealth of the rational investor is given by
\begin{equation}\label{wealthdef}
W_t = P_t  X_t + P_{ft} X_{ft}~ .
\end{equation}

The wealth of the rational investor changes from time $t$ to $t+1$ according to
\begin{equation}\label{wealthincr}
W_{t+1} - W_t = (P_{t+1} - P_t) X_t + (P_{ft+1}  - P_{ft})  X_{ft} + d_{t+1} X_t~.
\end{equation}
This expression takes into account that the wealth at time $t+1$ is determined
by the allocation choice at time $t$ and the new values of the risky and the risk-free asset
at time $t+1$, which includes the payment of the dividend
($W_{t+1} = P_{t+1} X_t + P_{ft+1} X_{ft} + d_{t+1} X_t$).
Let us introduce the variables
\begin{equation}\label{xtdef}
x_t := {P_t X_t \over W_t}~ , \qquad R_{t+1} := {P_{t+1} \over P_t}- 1~ , \qquad R_f := {P_{ft+1} \over P_{ft}} -1~ .
\end{equation}\
They are respectively the fraction $x_t $ of the rational investor's wealth invested
in the risky asset at time $t$, the discrete time return $R_{t+1}$ per stock of the risky asset
from time $t$ to $t+1$ and the risk-free rate of return $R_f$ assumed constant.
This allows us to rewrite \eqref{wealthincr} as giving the total relative wealth variation
from $t$ to $t+1$:
\begin{equation}
\label{wealthplone}
W_{t+1} - W_t  = W_t \left[ R_f + x_t \left(R_{t+1} - R_f + {d_{t+1} \over P_t}\right) \right] \equiv W_t \left[ R_f + x_t R_{\mathrm{excess},t+1} \right] ~ ,
\end{equation}
where we define 
\begin{equation}
R_{\mathrm{excess},t+1} = R_{t+1} - R_f + d_{t+1} / P_t
\label{Rexcessdef}
\end{equation}
as the excess return of capital and dividend gains over the risk-free rate.

The problem of the rational agent at time $t$ is to maximize the expected utility of his 
wealth for the next period by choosing the right proportion of wealth $x_t$ to invest in the risky asset, 
\begin{equation}
\mathrm{max}_{x_t}~ \E_t \left[ U(W_{t+1}) \right] ~ ,
\label{eqMaximization}
\end{equation}
where $\text{E}_t[.]$ means the expectation of the variable in the bracket performed
at time $t$, i.e., under the knowledge of available information up to and including time $t$.
If we assume the rational agent to have constant relative risk aversion, this proportion is constant in
time and wealth. This can be shown by employing the explicit utility function $U(W)$ exhibiting
constant relative risk aversion $\gamma$:
\begin{equation}
U(W)=\begin{cases}
  \log(W),  & \text{for }\gamma = 1~ ,\\
  \frac {W^{1-\gamma}}{1-\gamma}, & \text{for }\gamma \neq 1~ .
\end{cases}
\label{eqUtilities}
\end{equation}
Given this utility function and wealth evolution (\ref{wealthplone}), it is easy to see that
the maximization condition (\ref{eqMaximization}) is independent of $W_t$. 

We may obtain an approximate solution for $x_t$ in the special case where the 
wealth does not change much, i.e. in the case of small returns, so that the following expansion 
becomes approximately valid:
$R_f, R_{\mathrm{excess},t+1} \ll 1$.
\begin{equation}
\begin{split}
\E_t[U(W_{t+1})] & = U(W_t) + U'(W_t) W_t (R_f + x_t \E_t[R_{\mathrm{excess},t+1}])  \\ & + \frac 1 2 U''(W_t) W^2_t  x_t^2 \Var_t[ R_{\mathrm{excess},t+1}] + \mathcal{O}(R_f^3, R_{\mathrm{excess},t+1}^3)~ .
\end{split}
\label{eqEUseries}
\end{equation}
Maximizing this expression with respect to $x_t$ gives
\begin{equation}
x_t =  \frac {1}{\gamma} \frac {\E_t[R_{\mathrm{excess},t+1}]}{\Var_t[R_{\mathrm{excess},t+1}]}~ ,
\label{oteqtais}
\end{equation}
where
\begin{equation}
\gamma \equiv - \frac{W_t U''(W_t)}{U'(W_t)}~.
\end{equation}

In expression (\ref{oteqtais}), 
$\E_t[R_{\mathrm{excess},t+1}] \equiv \E_t[R_{t+1}]-R_f + {\text{E}_t[d_{t+1}] / P_t}$
represents the total excess expected rate of return of the risky asset
from time $t$ to $t+1$ above the risk-free rate.
In the following, we assume myopic rational agents who do
not learn but invest according to fundamental valuation. 
They expect a steady relative growth rate embodied by a constant total 
excess rate of return $R_{\rm excess}$, which is based on the behavior of stock
markets in the long run:
\begin{equation}
R_{\rm excess} := E_t[R_{t+1}]-R_f + {\text{E}_t[d_{t+1}] \over P_t}  = ~ {\rm constant}~.
\end{equation}
We will assume that $R_{\rm excess}>0$, so that the risky asset is desirable. The variance $\Var_t[R_{\mathrm{excess},t+1}]$
will be denoted by $\tilde \sigma^2$ and is given by
\begin{equation}
\tilde \sigma^2 := \Var_t[R_{\mathrm{excess},t+1}] =  \sigma^2 +  {\text{Var}[d_{t+1}] \over P_t^2}~ ,
 \qquad \sigma^2 := \text{Var}[R_{t+1}]~ .
\label{gjuyju}
\end{equation}
The expression for $\Var_t[R_{\mathrm{excess},t+1}]$ relies on the absence of correlation
between $R_{t+1}$ and $d_{t+1}$, because the dividend policy is assumed
independent of the market price and vice-versa, as in the dividend irrelevancy theory
of Modigliani and Miller (1958; 1963). Rational investors take the quantities $R_{t+1}$ and $d_{t+1}$
as exogenous to the price dynamics developed below, because they
reflect the information coming from a fundamental analysis.

In the sequel, 
we assume that ${\tilde \sigma}^2$ is constant, independent
of the price $P_t$. This corresponds to an expectation of the 
variance of dividends by the rational investors that tracks the 
(square of the) price. Alternatively, if 
$P_t  \gg \sqrt{ {\text{Var}[d_{t+1}] / \sigma^2}}$, 
${\tilde \sigma}^2 \simeq  \sigma^2$ and ${\tilde \sigma}^2$ is 
again approximately constant, as long as the rational investors
form a non-varying expectation of the volatility of future prices
of the risky asset. The assumption that ${\tilde \sigma}^2$ is constant
is also made by Chiarella et al. (2009) and in the framework of 
Boswijk et al. (2007),  if investors are assumed to be myopic, 
i.e. only look at the next period. 

Expression (\ref{oteqtais}) then becomes
\begin{equation}\label{oteqtais2gw}
x_t = x := {R_{\rm excess} \over \gamma {\tilde \sigma}^2} ~,
\end{equation}
which is a constant. Note that this is not an ad hoc assumption, but a consequence
of constant relative risk aversion and of the stationary nature of the dividend process.
In particular, because of the constant relative risk aversion 
of the rational investor, as already mentioned, $x$ is independent of the current wealth $W_t$ of the agent.
This allows us to treat all rational agents as one group with total wealth $W_t$ irrespective
of the distribution of the agents' individual wealth levels within the group. From here on,
we will call $W_t$ the wealth of \textit{the rational investors}.

The assumption, that the variance ${\tilde \sigma}^2$ given by (\ref{gjuyju}) is constant, implies
${\text{Var}[d_{t}] = ({\tilde \sigma}^2  - \sigma^2)  P_{t-1}^2}$. Therefore,
the flow of dividend $d_t$ follows the stochastic process
\begin{equation}
d_t = P_{t-1} \left[ r + \sigma_r   u_t \right]~,
\label{huj5iujewh}
\end{equation}
where $r := R_{\rm excess} - {\rm E}_{t-1}[R_t] + R_f$,
$\sigma_r = \sqrt{{\tilde \sigma}^2  - \sigma^2}$
and $u_t$ forms a series of standard i.i.d. random variables with distribution $N(0,1)$.

Thus, under the above assumptions, the rational
investors rebalance their portfolio so as to have a constant relative weight exposure
to the risky asset. This is equivalent to the 
traditional portfolio allocation benchmark of 70\% bonds and 30\% stocks used
by many mutual and pension funds. Rewriting expression (\ref{wealthincr})
with the condition of a fixed fraction $x$
invested in the risky asset, the wealth $W_t$ at time $t$
of the rational investors becomes at $t+1$
\begin{equation}
W_{t+1} = (P_{t+1} + d_{t+1}) x {W_t \over P_t} + (1-x) W_t (1+R_f)~.
\end{equation}

The excess demand of the risky asset from $t-1$ to $t$ of the group of rational investors is defined by
\begin{equation}\label{deldptxt}
\Delta D_t^\text{rational} := P_t X_t - P_t X_{t-1} = P_t X_t -{P_t \over P_{t-1}} ~ P_{t-1} X_{t-1}
= x W_t \left(1 - {P_t \over P_{t-1}} {W_{t-1} \over W_{t}} \right)~.
\end{equation} 
Expression (\ref{wealthincr}) with definitions (\ref{xtdef}) gives
\begin{equation}
{P_t \over P_{t-1}} {W_{t-1} \over W_{t}} = {P_t \over (P_t  + d_{t}) x + P_{t-1}(1-x) (1+R_f)}~.
\end{equation}
This allows us to rewrite the excess demand $\Delta D_t^\text{rational}$ as
\begin{equation}
\Delta D_t^\text{rational} = x W_{t-1} \left[ (1-x) {P_{t-1} (1+R_f) - P_t  \over P_{t-1}} + {x d_t \over P_{t-1}} \right]~,
\end{equation} 
where $x$ is given by expression (\ref{oteqtais2gw}). This last expression can be written, using (\ref{huj5iujewh}), as
\begin{equation}
\Delta D_t^\text{rational} = x W_{t-1} \left[ (1-x) {P_{t-1} (1+R_f) - P_t  \over P_{t-1}} + x (r + \sigma_r   u_t)\right]~.
\label{hyujik}
\end{equation} 
This corresponds to a kind of mean-reversing excess demand, where rational investors
tend to buy the risky asset when its price is low and vice-versa. 
But this mean-reversing excess demand is adjusted by taking into account two factors
that quantify an abnormal price increase (resp. decrease), which would justify unloading 
(resp. adding) the risky asset to the rational investors' portfolio. First, 
a price change is compared with the change that would occur if the corresponding wealth 
was instead invested in the risk-free asset. Second, even if its price 
decreases, the risky asset may still be attractive if it pays a sufficient dividend to compensate.

In absence of noise traders, the market clearing condition $\Delta D_t^\text{rational} =0$ leads to
\begin{equation}
P_t  = (1+R_f) P_{t-1}  + {x \over 1-x} d_t ~.
\label{whuri}
\end{equation} 
In the simplified case where the dividends $d_t$ are growing at
a constant rate $g>0$ such that $d_t = d_0 (1+g)^t$,
equation (\ref{whuri}) solves into
\begin{equation}
P_t  = (1+R_f)^{t} P_0 +  {x \over 1-x} (1+R_f)^{t} {d \over R_f-g} ~,
\end{equation} 
for $g < R_f$,  neglecting a term $\left[(1+g)/(1+R_f)\right]^t$ compared to $1$.
One recognizes the Gordon-Shapiro fundamental valuation, price $={\rm dividend}/(R_f-g)$, 
multiplied by a scaling factor taking into account the partitioning of the wealth of the rational investors
with the condition that a constant fraction is invested in the risky asset.

\subsection{Excess demand of the noise traders \label{hj6j8j}}

\subsubsection{General framework}

We assume that (a) the noise traders are characterized by polarized decisions
(in or out of the risky asset), (b) they tend to herd and (c) they are trend-followers.
A large body of literature indeed documents a lack-of-diversification puzzle
(Kelly, 1995; Baxter and Jermann, 1997; Statman, 2004) as
well as over-reactions (Werner et al., 1987; 1990; Frank, 2004).
There is strong evidence for imitation and herding, even among sophisticated
mutual fund managers (Wermers, 1999), and technical analysis and chart trading is ubiquitous.

We account for the observations of lack-of-diversification by assuming that a noise
trader is fully invested either in the risky asset or in the risk-free asset. 
In contrast to the rational agents, our noise traders have different opinions,
which fluctuate stochastically according to laws given below. The number of
noise investors invested in the risky asset (respectively invested
in the risk-free asset) is $N^+_t$ (respectively $N^-_t$), and we have
\begin{equation}
N^+_t+ N^-_t \equiv N_\text{noise}~ .
\label{jrujwrb2}
\end{equation}

We do not aim at describing the heterogeneity between noise traders, which has been shown to lead
to fat-tailed distribution of their wealth as a result of heterogenous investment
decisions (Bouchaud and Mezard, 2000; Klass et al., 2007;
Harras and Sornette, 2011). This is not a restriction
in so far as we consider their aggregate impact. 

Therefore, as for the rational investors, we treat the noise traders as one group with total wealth $W^n_t$.
The ratio of wealth of the group of noise traders invested in the risky asset corresponds to the ratio of 
bullish investors among the population of noise traders. Let us denote this quantity at time $t$ by
\begin{equation}
\label{eqXnt}
x^n_t := \frac {N^+_t}{N_\text{noise} }~.
\end{equation}
Then, the wealth $W^n_t$ of noise traders at time $t$ becomes at $t+1$
\begin{equation}
W^n_{t+1} = (P_{t+1} + d_{t+1}) x^n_t {W^n_t \over P_t} + (1-x^n_t) W^n_t (1+R_f)~.
\label{huyiu}
\end{equation}
The excess demand of the noise traders over the time interval $(t-1,t)$ is equal to
\begin{equation}
\Delta D_t^\text{noise} =  x^n_t W^n_t - \frac{P_t}{P_{t-1}} x^n_{t-1} W^n_{t-1} = 
\end{equation}
\begin{equation}\label{resodjdd}
W^n_{t-1} \left[ x^n_t(1-x^n_{t-1})(1+R_f) - x^n_{t-1}(1-x^n_t)\frac {P_t}{P_{t-1}} + x^n_t x^n_{t-1}\frac{d_t}{P_{t-1}} \right].
\end{equation}

Let us introduce the opinion index (Lux and Marchesi, 1999)
\begin{equation}
s_t := {N^+_t-N^-_t\over N_\text{noise}} \in [-1,1] ~,
\label{jrunbwfv}
\end{equation}
which can be interpreted as the aggregate bullish ($s_t >0$) versus
bearish  ($s_t <0$) stance of the noise traders with respect to the risky asset.
With this definition (\ref{jrunbwfv}) and with (\ref{jrujwrb2}), we have
\begin{equation}
{N^+_t \over N_\text{noise}} = {1 \over2} (1+s_t) = x^n_t , \quad {N^-_t \over N_\text{noise}} = {1 \over2} (1-s_t) = 1-x^n_t ~ .
\label{jeyjyb2}
\end{equation}
Expression (\ref{resodjdd}) with (\ref{jeyjyb2}) yields
\begin{equation}
\label{jnbgw}
\Delta D_t^\text{noise}  = \frac{W^n_{t-1}}{4 P_{t-1}} \left[ (1+s_t)(1-s_{t-1})(1+R_f)P_{t-1} - (1-s_t)
(1+s_{t-1}) P_t + (1+s_t)(1+ s_{t-1})d_t \right]~.
\end{equation}

\subsubsection{Master equation for the bullish/bearish noise trader unbalance $s_t$}

Let us now specify the dynamics of the opinion index $s_t$. We assume that, at each time step,
each noise trader may change her mind and either sell her risky portfolio if 
she was previously invested or buy the risky portfolio if she had only the risk-free asset.
Again, we assume an all-or-nothing strategy for each noise trader at each time step.
Let $p^+_{t-1}$ be the probability that any of the $N^+_{t-1}$ noise traders
who is currently fully invested in the risky portfolio decides 
to remove her exposure during the time interval $(t-1,t)$. 
Analogously, let $p^-_{t-1}$ be the probability that any of the 
$N^-_{t-1}$ traders who are currently (at time $t-1$) out of the risky market
decides to buy it. For a noise trader $k$ who owns the risky asset, 
her specific decision is represented by the random variable $\zeta_k(p^+)$,
which takes the value $1$ (sell) with probability $p^+$ and the value $0$
(keep the position) with probability $1-p^+$.
Similarly, for a noise trader $j$ who does not own the risky asset, 
her specific decision is represented by the random variable $\xi_j(p^-)$,
which takes the value $1$ (buy) with probability $p^-$ and the value $0$ 
(remain invested in the risk-free asset) with probability $1-p^-$.
For given $p^+$ and $p^-$, 
the variables $\{\xi_j(p^+)\}$ and $\{\zeta_k(p^-)\}$ are i.i.d.. 

Aggregating these decisions over all noise traders invested in the risky asset at time $t$, we have
\begin{equation}
N^+_t =  \sum_{k=1}^{N^+_{t-1}} [1-  \zeta_k (p^+_{t-1})] + \sum_{j=1}^{N^-_{t-1}} \xi_j (p^-_{t-1}) ~.
\label{hhuj5u4y}
\end{equation}\
The first term in the r.h.s. of (\ref{hhuj5u4y}) corresponds to all the traders who held
the risky asset at $t-1$ and continue to hold it at $t$. The second term in the r.h.s. of (\ref{hhuj5u4y}) 
represents the noise traders who were holding the risk-free asset at $t-1$ and sold it
to buy the risky asset at time $t$. Similarly, 
\begin{equation}
N^-_t =  \sum_{k=1}^{N^+_{t-1}}  \zeta_k (p^+_{t-1}) + \sum_{j=1}^{N^-_{t-1}} [1-\xi_j (p^-_{t-1})] ~.
\label{hhuj5awau4y}
\end{equation}
The opinion index $s_t$ (\ref{jrunbwfv}) is thus given by
\begin{equation}
s_t = {1 \over N_\text{noise}} \left( \sum_{k=1}^{N^+_{t-1}} [1- 2 \zeta_k (p^+_{t-1})] + \sum_{j=1}^{N^-_{t-1}} 
[2 \xi_j (p^-_{t-1}) -1] \right)~.
\label{jrunrttbwfv}
\end{equation}
Using the i.i.d. property of the $\{ \xi_j(p) \}$ and $\{ \zeta_k(p) \}$ variables allows us to obtain the following
exact expression for the mean of $s_t$:
\begin{equation}
\label{etameanvar}
\text{E}\left[s_t\right] = s_{t-1} + p^-_{t-1}(1- s_{t-1}) - p^+_{t-1}(1+s_{t-1})~.
\end{equation}

\subsubsection{Influence of herding and momentum on the behavior of noise traders}

As can be seen from (\ref{jnbgw}) together with (\ref{jrunrttbwfv}),
the probabilities $p^\pm$ embody completely the behavior of the noise traders.
We assume that $p^\pm$ at time $t-1$ are both a function of $s_{t-1}$ (social imitation effect)
defined by (\ref{jrunbwfv}) and of a measure $H_t$  of the price momentum given by
\begin{equation}
H_t = \theta H_{t-1} + (1-\theta) \left({P_t \over P_{t-1}} - 1\right)~,
\label{wh3u6juj}
\end{equation}
which is nothing but the expression for an exponential moving average of the history of past returns.
The parameter $0 \leq \theta <1$ controls the length of the memory 
that noise traders keep of past returns, the closer to $1$, the longer the memory $\sim 1/(1-\theta)$. 

Considering that the probabilities $p^\pm$ are functions of $s_{t-1}$ and $H_{t-1}$,
\begin{equation}
p^\pm_{t-1} = p^\pm(s_{t-1},H_{t-1}) ~,
\end{equation}
means that the noise traders make their decisions to buy or sell the risky Internet stock
based on (i) the majority view held by their group and (ii) the recent capital gains
that the risky asset has provided over a time frame $\sim 1/(1-\theta)$.
We assume the noise traders buy and sell symmetrically with no bias:
a strong herding  in favor of the risky asset or a strong positive momentum
has the same relative effect on the drive to buy (or to sell) than a strong negative sentiment
or strong negative momentum on the push to sell (or to buy).  This is expressed 
by the following symmetry relation
\begin{equation}
p^-(s,H) = p^+(-s,-H) ~.
\label{hwryyhwtb2t}
\end{equation}
The simplest functions satisfying (\ref{hwryyhwtb2t}) are the linear expressions\footnote{Another possibility that is not 
further explored in this 
paper consists of employing the hyperbolic tangent: $p^\pm(s,H) = \frac 1 2 \left[ 1 \mp \kappa / p \tanh(s + H) \right]$. 
These transition probabilities correspond to the Glauber transition rates of an ensemble of spins on a fully 
connected graph with equal interaction strengths, see for example Harras et al. (2012).}
\begin{equation}\label{rhopmexpr}
p^-(s,H) = {1\over2}\left[p + \kappa \cdot (s + H)\right] ~, \quad p^+(s,H) = {1\over2}\left[p - \kappa\cdot (s + H)\right] ~.
\end{equation}
This defines two parameters $p$ and $\kappa$, chosen sufficiently small such that 
$p^-(s,H)$ and $p^+(s,H)$ remain between $0$ and $1$. 
The positive parameter $p$ controls the average holding time of the positions in the absence of any other influence.
In other words, a position will last typically $\sim 2/p$ time steps in the absence
of social imitation and momentum influence. The parameter $\kappa$ quantifies
the strength of social interactions and of momentum trading. 
Instead of $\kappa$, one could use two 
parameters for the opinion index and momentum, respectively. For the sake of parsimony we will only work with one 
parameter treating $s$ and $H$ symmetrically.
For instance, for  $\kappa>0$,
if there is already a majority of agents holding the risky asset and/or if its price
has been increasing recently, then the probability for noise traders holding the risk-free asset
to shift to the risky asset is increased and the probability for the noise traders
who are already invested to sell their risky asset is decreased. The reverse holds 
for  $\kappa <0$, which describes ``contrarian'' traders.  In the sequel, 
we will only consider the case  $\kappa>0$, which describes imitative and
trend-following agents. Generalizations to allow for additional heterogeneous
beliefs, involving mixtures as well as adaptive imitative and contrarian agents,
is left for other communications. In this spirit, let us mention that Corcos et al. (2002) have
introduced a simple model of imitative agents who turn contrarian
when the proportion of herding agents is too large, which generates chaotic
price dynamics.

Putting expressions (\ref{rhopmexpr}) in (\ref{etameanvar}) yields
\begin{equation}
\label{etameargrtanvar}
\text{E}\left[s_t\right] = (1 + \kappa - p) s_{t-1} + \kappa H_{t-1}~.
\end{equation}

\section{Dynamical market equations}

\subsection{Market clearing condition and price dynamics}

The equation for the risky asset price dynamics is obtained from 
the condition that, in the absence of external supply,
the total excess demand summed over the rational and the noise traders vanishes:
\begin{equation}
\Delta D_t^\text{rational} + \Delta D_t^\text{noise} = 0 ~.
\label{fhetyjbv}
\end{equation}
In other words, the net buy orders of noise traders are satisfied by the net sell orders of the
rational traders, and vice-versa. 
Substituting in (\ref{fhetyjbv}) expression \eqref{hyujik} for the
excess demands $\Delta D_t^\text{rational}$ of the rational investors
and equation  \eqref{jnbgw} for the excess demand $\Delta D_t^\text{noise}$
of the noise traders, we obtain the price equation
\begin{eqnarray}
\frac {P_t}{P_{t-1}} & = & \left[ (1+s_t)\left((1+R_f)(1-s_{t-1})+(r+\sigma_r u_t)(1+s_{t-1}) \right) W^n_{t-1}  \right. \nonumber \\ 
\label{mclcondtransf} && \left. + \, 4 x \left( (1+R_f)(1-x) + (r + \sigma_r u_t) x \right) W_{t-1}  \right] / \\
&& \left[ (1+s_{t-1})(1-s_t) W^n_{t-1} + 4 W_{t-1} x(1-x) \right]. \nonumber 
\end{eqnarray}
Expression (\ref{mclcondtransf}) shows that the price of the risk asset
changes as a result of two stochastic driving forces: 
(i) the dividend-price ratio $(r+\sigma_r u_t)$ and (ii) the time increments of 
the bullish/bearish noise trader unbalance $\{s_t\}$.
The impact of $\{s_t\}$ is controlled by the wealth of the group of noise traders $W^n_{t-1}$. 
As we shall demonstrate below, 
this becomes particularly important during a bubble where trend-following noise traders 
tend to gain much more than rational investors. With the increasing influence of noise traders,
the market becomes much more prone to self-fulfilling prophecies. Rational traders are less
able to attenuate the irrational exuberance -- they simply do not have enough wealth
invested in the game.

\subsection{Complete set of dynamical equations}

Let us put all ingredients of our model together to state concisely all the 
equations controlling the price dynamics coupled with the opinion forming
process of the noise traders. 
As discussed above, the wealth levels of the rational and noise traders
are also time-dependent and influence the market dynamics. We thus arrive at the following
equations.

\vskip 0.3cm
\noindent
{\it Dynamics of the noise traders opinion index}:
\begin{equation}
\label{eqOpinionIndex}
s_t = \frac 1 {N_\text{noise}} \left( \sum_{k=1}^{N_\text{noise}(1+s_{t-1})/2}\left[ 1-2 \zeta_k(p^+_{t-1})\right] + \sum_{j=1}^{N_\text{noise}(1-s_{t-1})/2}\left[2\xi_j(p^-_{t-1}) - 1\right] \right)~,
\end{equation}
where $\zeta_k(p^+_{t-1})$ takes the value $1$ with probability $p^+_{t-1}$ and the value $0$
with probability $1-p^+_{t-1}$, $\xi_j(p^-_{t-1})$  takes the value 
$1$ with probability $p^-_{t-1}$ and the value $0$ with probability $1-p^-_{t-1}$, and
$p^+_{t-1}$ and $p^-_{t-1}$ are given by expressions (\ref{rhopmexpr}):
\begin{equation}
\label{rhopmerrtexpr}
\begin{split}
p^-_{t-1}(s_{t-1},H_{t-1}) = & {1\over2}\left[p + \kappa \cdot (s_{t-1} + H_{t-1})\right] ~, \\ 
\quad p^+_{t-1}(s_{t-1},H_{t-1}) = & {1\over2}\left[p - \kappa\cdot (s_{t-1} + H_{t-1})\right] ~.
\end{split}
\end{equation}
Thus, $\text{E}\left[ s_t\right]$ given by expression (\ref{etameargrtanvar}).

\vskip 0.3cm
\noindent
{\it Dynamics of the risky asset price}:
\begin{eqnarray}\nonumber
P_t /  P_{t-1}  & = & \left[ (1+s_t)\left((1+R_f)(1-s_{t-1})+(r+\sigma_r u_t)(1+s_{t-1}) \right) W^n_{t-1}  + \right. \\
\label{eqPrice} && \left. + \, 4 x \left( (1+R_f)(1-x) + (r + \sigma_r u_t) x \right) W_{t-1}  \right] /   \label{heywgw} \\
&& \left[ (1+s_{t-1})(1-s_t) W^n_{t-1} + 4 x(1-x) W_{t-1} \right] ~.\nonumber
\end{eqnarray}

\vskip 0.3cm
\noindent
{\it Weath dynamics of rational investors}:
\begin{equation}
\label{eqWealthEvolutionRational}
W_t / W_{t-1} =  x \left(\frac {P_t }{P_{t-1}} + (r+\sigma_r u_t) \right) + (1-x)(1+R_f)~~.
\end{equation}

\vskip 0.3cm
\noindent
{\it Weath dynamics of noise traders}:
\begin{equation}
\label{eqWealthEvolutionNoise}
W^n_t / W^n_{t-1} = \frac {1+s_{t-1}} 2 \left(\frac {P_t }{P_{t-1}} + (r+\sigma_r u_t) \right) + \frac {1-s_{t-1}} 2(1+R_f)~.
\end{equation}

\vskip 0.3cm
\noindent
{\it Momentum of the risky asset price}:
\begin{equation}
\label{eqMomentum}
H_t = \theta H_{t-1} + (1-\theta) \left( \frac {P_t}{P_{t-1}} - 1 \right)~.
\end{equation}
And $u_t$ forms a series of standard i.i.d. random variables with distribution $N(0,1)$.

The set of equations (\ref{eqOpinionIndex}) to (\ref{eqMomentum}) together 
with the realization of the stochastic dividend process $u_t$ completely specify the model and its dynamics.
Equation (\ref{eqOpinionIndex}) describes how noise traders form 
their opinion $s_t$ based on the previous prevalent opinion
$s_{t-1}$ and the recent price trend $H_t$. Rational traders stick to their choice of investing $x$ in the risky asset.
Equation (\ref{heywgw}) gives the new market price $P_t$ when excess demands of both groups are matched.
Equations (\ref{eqWealthEvolutionRational}) and (\ref{eqWealthEvolutionNoise})
 describe the evolution of the wealth levels $W_t$ and $W^n_t$
 for rational and noise traders, respectively. There are capital gains and dividend gains from the risky asset,
 and interest payments by the risk-free asset. 
 The new market price also feeds into the momentum of the risky asset described by equation
 (\ref{eqMomentum}).

We have the following flow of causal influences:
\begin{enumerate}
\item The recent price trend $H_{t-1}$ and the prevailing opinion $s_{t-1}$ among noise traders determine the 
investment decision of noise traders governed by $s_t$, while rational traders invest a constant fraction $x$ of their wealth.

\item Market clearing determines the price $P_t$ based on investment decisions $x$ and $s_t$, and 
previous wealth levels $W_{t-1}$ and $W^n_{t-1}$ for rational and noise traders, respectively.

\item The new wealth levels $W_{t}$ and $W^n_{t}$ are based on the market price $P_t$ and
investment decisions $x$ and $s_t$.

\end{enumerate}

\subsection{Control parameters and their time-scale dependence \label{hjyjue}}

The set of equations (\ref{eqOpinionIndex}) to (\ref{eqMomentum})
 depends on the following parameters:
\begin{enumerate}
\item $x$ quantifies the constant fraction of wealth that rational traders invest in the risky asset. 
\item $\theta$ fixes the time scale over which noise traders estimate price momentum. By construction,
$0 \leq \theta <1$.
\item $N_{\rm Noise}$ is the number of noise traders that controls the fluctuations
of the majority opinion of noise traders.
\item $p$ controls the average holding time of the positions of noise traders in the absence of any other influence.
\item $\kappa$ quantifies the strength of social interactions and of momentum trading
by noise traders. 
\item $R_f$ is the rate of return of the risk-free asset.
\item $r$ and $\sigma_r$ are the mean and standard deviation of the dividend-price ratio.
\end{enumerate}

In order to have an intuitive understanding of the role and size of these parameters,
it is useful to discuss how they depend on the time scale over which traders
reassess their positions.  Until now, we have expressed the time $t$ in units of a unit step $1$,
which could be taken for instance to be associated with the circadian rhythm,  i.e., one day.
But there is no fundamental reason for this choice and our theory has the same formulation
under a change of the time step. Let us call $\tau$ the  time interval between 
successive reassessments of the rational  investors, with
$\tau$ being measured in a calendar time scale, for instance, in seconds, hours or days.

First, the parameters $N_\text{noise}$ and $N_\text{rational}$ are a priori independent of $\tau$, while
they may be a function of time $t$. We neglect this dependence as we are interested
in the dynamics over time scales of a few years that are characteristic of bubble regimes.
The parameter $\gamma$ is also independent of $\tau$.

In contrast, the parameters $R_f$, $r$ and $\sigma_r^2$ are functions of $\tau$,
as the return of the risk-free asset, the average expected dividend return and its variance 
depend on the time scale. The simplest and standard dependence of 
Wiener processes or discrete random walks is $R_f \sim r \sim \sigma_r^2 \sim \tau$.
Because of its definition, 
$x = {R_{\rm excess} / \gamma {\tilde \sigma}^2}$, the fraction
of wealth $x$ rational investors hold is independent of time.

By construction, the parameter $\theta$ characterizing the memory of the 
price momentum influencing the decisions of noise traders depends on $\tau$.
This can be seen by replacing $t-1$ by $t-\tau$ to make explicit the unit time scale 
in expression (\ref{wh3u6juj}), giving
\begin{equation}
{H_t-H_{t-\tau}\over \tau} = {1-\theta\over \tau} \left( {P_t \over P_{t-\tau}} - 1 - H_{t-\tau}\right)~ .
\end{equation}
Requesting a bona-fide limit for small $\tau$'s leads to 
\begin{equation}
{1-\theta\over \tau} = \varrho = \text{const}~ ,
\label{jhwb2}
\end{equation}
where the time scale $ \mathcal{T}_H := 1/\varrho$ is the true momentum memory. Thus, we have
\begin{equation}
\label{oneminthvarrho}
1-\theta = \varrho \cdot \tau , \qquad \mathcal{T}_H := {1\over \varrho} = {\tau\over 1-\theta} ~.
\end{equation}

\section{Theoretical analysis and super-exponential bubbles \label{heytbrgr}}

\subsection{Reduction to deterministic equations}

It is possible to get an analytical understanding
of the solutions of the set of equations   (\ref{eqOpinionIndex}) to (\ref{eqMomentum})
if we reduce them into their deterministic components.
The full set including their stochastic contributions will be studied with 
the help of numerical simulations in the next section.

Taking $u_t\equiv 0$ and replacing $s_t$ by its expectation 
$\text{E}[s_t]$ given by \eqref{etameargrtanvar}, we obtain the following 
deterministic equations
\vskip 0.2cm
\noindent
{\it Dynamics of the noise traders opinion index}:
\begin{equation}
\label{eqDetS}
s_t = (1+\kappa - p) s_{t-1} + \kappa H_{t-1}~,
\end{equation}
{\it Dynamics of the risky asset price}:
\begin{eqnarray}\nonumber
P_t /  P_{t-1}  & = & \left[ (1+s_t)\left((1+R_f)(1-s_{t-1})+r (1+s_{t-1}) \right) W^n_{t-1}  + \right. \\
&& \label{eqDetP} \left. + \, 4 x \left( (1+R_f)(1-x) + r x \right) W_{t-1}  \right] / \\
&& \left[ (1+s_{t-1})(1-s_t) W^n_{t-1} + 4 x(1-x) W_{t-1}  \right].\nonumber
\end{eqnarray}
{\it Weath dynamics of rational investors}:
\begin{equation}
\label{eqDetW}
W_t / W_{t-1} =  x \left(\frac {P_t }{P_{t-1}} + r \right) + (1-x)(1+R_f)~,
\end{equation}
{\it Weath dynamics of noise traders}:
\begin{equation}
\label{eqDetWn}
W^n_t / W^n_{t-1} = \frac {1+s_{t-1}} 2 \left(\frac {P_t }{P_{t-1}} + r \right) + \frac {1-s_{t-1}} 2(1+R_f)~,
\end{equation}
{\it Momentum of the risky asset price}:
\begin{equation}
\label{eqDetH}
H_t = \theta H_{t-1} + (1-\theta) \left( \frac {P_t}{P_{t-1}} - 1 \right)~.
\end{equation}

This system of five coupled deterministic equations is non-linear and 
completely coupled, there is no autonomous
subsystem. In particular, the multiplicative price equation is highly non-linear.
The wealth equations describe the multiplicative process of capital accumulation depending
on the choice of how to split the portfolio on the risky and risk-free asset yielding capital gains,
given the dividend gains and the risk-free rate. 

\subsection{Fixed points and stability analysis}

\label{subsecFixed}

To gain insights into the system of coupled equations, we will consider the scenarios in which the wealth levels of rational and noise traders grow at
the same steady rate $\rho$,
\begin{equation}
W_t = W_0 \rho^t~, \quad W^n_t = W^n_0 \rho^t~.
\end{equation}
These scenarios are those for which the two populations of rational and noise
traders remain relevant in the economy. In other cases in which
one of the two populations sees his wealth growing faster than the other one, at long times, it will completely
dominate the economy, leading to unrealistic and trivial dynamics.

This allows us to decouple the equations for $H_t$, $s_t$ and $P_t$ from the wealth 
equations. Assuming equal initial wealth levels $W_0^n = W_ 0$, 
the fixed points $\{ (H^*, s^*)\}$ are determined by the system:
\begin{equation}
H^* = R_f + r \frac {(1+s^*)^2 + 4 x^2}{(1+s^*)(1-s^*) + 4 x(1-x)}~,
\end{equation}
\begin{equation}
s^* = \frac \kappa {p - \kappa} H^*~.
\end{equation}
Since this system is essentially one third-order equation, it can be solved analytically
yielding three fixed points. As we will see later, for typical parameter values, there
is one solution $s^*, H^* \ll 1$, while the other two lie outside the restricted domain
of $[-1,1]$ for $s$. It can be estimated as
\begin{equation}
H^* = R_f + \frac {1+4x}{1+4x-4x^2} r + \mathcal O (r^2, R_f^2)~,
\end{equation}
\begin{equation}
s^* = \frac \kappa {p-\kappa} \left[R_f + \frac {1+4x}{1+4x-4x^2} r  + \mathcal O (r^2, R_f^2) \right]~.
\end{equation}
This fixed point is stable for $\kappa < p$  over a 
range of the other parameter values and is unstable for $\kappa > p$. A deviation from the fixed point
due to stochastic fluctuations in the opinion index leads to a price change in the same
direction. According to (\ref{eqDetS}), for $\kappa > p$, the opinion index grows transiently
exponentially (until its saturation). 
Since the stability is mainly governed by the relative value of the two parameters 
$\kappa$ and $p$ characterizing noise trader behavior, we conclude  
that there is an inherent instability caused by herding and trend following,
which is independent of the stochastic dividend process.

\subsection{Super-exponential bubbles}

It is well-known that many bubbles in financial markets 
start with a phase of super-exponential growth, see 
for example Sornette et al. (2009) for oil prices, Jiang et al. (2010) for the Chinese stock market
and Yan et al. (2012) for major equity markets. Furthermore, 
Sornette et al. (2013) discuss various theoretical and empirical questions related to
faster-than-exponential growth of asset prices,  
while H\"usler et al. (2013) document super-exponential bubbles in a controlled experiment in the 
laboratory.

Phases with faster-than-exponential growth of the price are inherent also in the present model.
If a bubble is essentially driven by herding and trend following, we may neglect
the dividend process and expand the pricing formula (\ref{eqDetP}) in terms of $r$ and $R_f$:
\begin{equation}
\frac {P_t}{P_{t-1}}  = \frac {(1+s_t)(1-s_{t-1}) + 4 x (1-x) W_{t-1}/W^n_{t-1}}{(1-s_t)(1+s_{t-1}) + 
4 x (1-x) W_{t-1}/W^n_{t-1}} + \mathcal O (r, R_f)~.
\end{equation}
Again, we use focus on the scenario that the ratio of the wealth levels $W_{t-1}$ and $W^n_{t-1}$ 
of the rational and noise traders remains approximately constant,
\begin{equation}
W_{t-1}/W^n_{t-1} = W_{t_0} / W^n_{t_0} = \mathrm{const}~.
\end{equation}
This is the case if both wealths grow at the same constant exponential growth rate
or, more accurate here, if they both grow super-exponentially in the same way.
Starting with an opinion index $s_0$ at time $t = t_0$, we can further simplify the price 
equation to:
\begin{equation}
\frac {P_t}{P_{t-1}} = 1 + b (s_t - s_{t-1}) + \mathcal O (r, R_f, (s-s_0)^2)~,
\end{equation}
where the constant quantity $b$ is of order 1 provided the initial levels of wealth are of 
the same order of magnitude: 
\begin{equation}
b =  \frac 2 {1+4 x (1-x) W_{t_0}/W^n_{t_0} - s_0^2} \sim \mathcal O (1)~.
\end{equation}
Therefore, up to terms of order $\mathcal O (r, R_f, (s-s_0)^2)$, the price evolves as
\begin{equation}
\label{eqExpP}
\frac {P_t}{P_0} = \prod_{j=1}^t \left[1 + b (s_j - s_{j-1})\right] \simeq \prod_{j=1}^t \mathrm e^{b (s_j - s_{j-1})} = \mathrm e^{b(s_t - s_0)}~.
\end{equation}
Since $s_t$ grows exponentially with time according to expression (\ref{eqDetS}) for $\kappa > p$,
the price $P_t$ grows as an exponential of an exponential of time. In other words, 
for the regimes when the opinion index grows exponentially ($\kappa > p$), we expect super-exponential 
bubbles in the price time series. Since our equations are symmetric in the sign of the opinion index
$s_t$, the same mechanism leads also to ``negative bubbles''
for a negative herding associated with a transition from bullish to bearish behavior
for which the price drops also super-exponentially in some cases.

\subsection{Time-dependent social impact and bubble dynamics}

The strength of herding is arguably regime dependent. In some phases, noise traders
are prone to herding, while at other times, they are more incoherently disorganized
``noise'' traders. This captures in our dynamical framework the phenomenon
of regime switching (Hamilton, 1989; Lux, 1995; Hamilton and Raj, 2002; Yukalov et al., 2009;
Binder and Gross, 2013; Fisher and Seidl, 2013; Kadilli, 2013), where successive phases 
are characterized by changing
values of the herding propensity. In this respect, we follow the model approach 
of Harras et al. (2012) developed in a similar context
and assume that the strength $\kappa$ of social interactions
and momentum influence slowly varies in time. In this way, we incorporate the effects of a changing world
on financial markets such as a varying economic and geopolitical climate into the model. More generally,
we allow for varying uncertainties influencing the behavior of noise traders. 
As we shall show, this roots the
existence of the bubbles documented below in the mechanism of ``sweeping of an instability''
(Sornette, 1994; Stauffer and Sornette, 1999).

More specifically, we propose that $\kappa$ undergoes a discretized 
Ornstein-Uhlenbeck process:\footnote{Choosing a confined random walk yields similar results, but 
the mean reversion is then effectively nonlinear (or threshold based), which is less standard.}
\begin{equation}
\label{eqKappaOU}
\kappa_t - \kappa_{t-1} = \eta (\mu_\kappa - \kappa_{t-1}) + \sigma_\kappa v_t~.
\end{equation}
Here $\eta > 0$ is the mean reversion rate, $\mu_\kappa$ is the mean reversion level
and $\sigma_\kappa > 0$ is the step size of the 
Wiener process realized by the series $v_t$ of standard i.i.d. random variables with
distribution $N(0,1)$. 

Our approach is related to how Lux (1995) describes switching between bear and bull markets.
While we propose a stochastic process for the strength of social interactions $\kappa$, Lux
adds a new deterministic term proportional to $\mathrm d \log P_t / \mathrm d t$ 
to the transition probabilities, which corresponds to a direct positive feedback.

The interesting case is $\mu_\kappa \lesssim p$, where $\kappa$ is on average
below the critical value $p$ but, due to stochastic fluctuations, 
may occasionally enter the regime with faster-than-exponential
growth $\kappa > p$ described  in the previous subsection.
Since an Ornstein-Uhlenbeck process with deterministic
initial value is a Gaussian process, its distribution is fully determined by the first and second moments.
Starting from an initial value $\kappa_0$, the non-stationary mean and covariance are given by:
\begin{equation}
\label{eqOUmean}
\mathrm E[\kappa_t] = \kappa_0 \mathrm e^{-\eta t} + \mu_\kappa \left(1 - \mathrm e^{-\eta t}\right)~,
\end{equation}
\begin{equation}
\mathrm{Cov}[\kappa_s,\kappa_t] = \frac {\sigma_\kappa^2}{2 \eta} \left( \mathrm e^{-\eta (t-s)} + \mathrm e^{-\eta (t+s)} \right)~, \quad s<t~.
\end{equation}
Both moments converge such that in the long run $\kappa_t$ admits the following 
stationary distribution:
\begin{equation}
\label{eqOUstationaryDistribution}
\kappa_t \sim N\left(\mu, \frac {\sigma_\kappa}{\sqrt{2 \eta}}\right)~.
\end{equation}
If, at some time $t$, the social interaction strength is above the critical value 
$\kappa_t \equiv \kappa_0 > p$, 
the time $\Delta T$ needed for $\kappa_t$ to revert 
to the subcritical regime $\kappa_t < p$ can be estimated from equation (\ref{eqOUmean}): 
\begin{equation}
\label{eqDeltaT}
\Delta T = \frac 1 \eta \log\left(\frac{\kappa_0 - \mu_\kappa}{p-\mu_\kappa}\right)~.
\end{equation}
Expressions (\ref{eqOUstationaryDistribution}) and (\ref{eqDeltaT}) will allow us 
to estimate how often the group of noise traders will interact in the supercritical regime
of the opinion index related to transient faster-than-exponential growth in the price 
 and how long a typical bubble will last.

\section{Numerical simulations and comparison with the dotcom bubble}

\subsection{Estimation of parameter values}

Let us take $\tau=1$ day and assume a typical memory used by noise traders
for the estimation of price momentum 
equal to about one month. This amounts approximately to $20$ trading days, hence
$\mathcal{T}_H \simeq {\tau \over 1-\theta} = 20$, leading to $\theta=0.95$.

We calibrate the average dividend-price ratio $r$ and its standard deviation
$\sigma_r$ to the values given by Engsted and Pedersen (2010), which are quite similar for 
various countries. We set the mean daily dividend-price ratio to $r = 1.6 \cdot 10^{-4}$ 
and the daily standard deviation to $\sigma_r = 9.5 \cdot 10^{-4}$. Furthermore, we assume a 
constant return of the risk-free asset of annualized $2\%$, i.e. a daily value of $R_f = 8 \cdot 10^{-5}$.

Rational investors keep $30\%$ of their wealth in the risky asset, that is, $x = 0.3$. The
wealth levels $W_t$ and $W^n_t$ of rational and noise traders 
evolve dynamically and determine the relative 
influence of the two groups. We analyze the importance of the initial endowments
$W_0$ and $W^n_0$ on the stability of the market. We capture this by the parameter
$\nu = W^n_0 / W_0$ and set $\nu$ to $1$, $1/2$ or $2$ in three
different sets of simulations.\footnote{Note that this is 
equivalent to setting the ratio of group sizes $\nu = N_\text{rational} / N_\text{noise}$ with the
assumption that both groups consist of representative agents with
equal initial wealth. In our formulation, $N_\text{noise}$
has no further importance than controlling the randomness of the opinion index. Thus
it disappears from the deterministic equations (\ref{eqDetS}) to (\ref{eqDetH}). }

For the parameter $p$ entering in expressions \eqref{rhopmexpr}, recall that it is equal to 
twice the probability that during a given day some noise trader will buy (or sell) the risky asset.
We posit $p=0.2$, which means that the natural trading frequency of traders in absence of social
influence is about two weeks. For the parameter $\kappa$ in \eqref{rhopmexpr} describing 
the strength of social interactions and of momentum trading, we assume that it is close to 
the parameter $p$. Specifically, for the Ornstein-Uhlenbeck process given in expression
(\ref{eqKappaOU}), we choose $\mu_\kappa= 0.98 p = 0.196$. We set the mean reversion speed 
$\eta$ and the step size $\sigma_\kappa$ such that (i) the Ornstein-Uhlenbeck process has a 
standard deviation of $0.1 p$ and (ii) a deviation of $\kappa_t$ two standard deviations
above $\mu_\kappa$ in the supercritical regime will revert within $\Delta T = \mathcal{T}_H = 20$:
\begin{equation}
\eta = \frac 1 {\Delta T} \log\left(\frac{\mu_\kappa + 2 \cdot 0.1 p - \mu_\kappa}{p-\mu_\kappa}\right) 
= \log(10)/20 \simeq 0.11 ~,
\end{equation}
\begin{equation}
\sigma_\kappa = 0.1p \sqrt{2 \eta} \simeq 0.001 ~.
\end{equation}
Summarizing, the numerical simulations presented in the figures correspond to
\begin{equation}\label{parameters}
\theta = 0.95 , \quad r = 1.6 \cdot 10^{-4} , \quad \sigma_r = 9.5 \cdot 10^{-4} , \quad R_f = 8 \cdot 10^{-5} , \quad x = 0.3 ~,
\end{equation}
\begin{equation}\label{parameters2} 
p=0.2 , \quad \mu_\kappa = 0.196 , \quad \sigma_\kappa = 0.001 , \quad \eta = 0.11 ~,
\end{equation}
and $\nu$ will be varied as $\nu = 0.5, 1, 2$. Furthermore, we run the simulations over
20 trading years, i.e. $T = 5000$.
 
We can now test our claims from the fixed points analysis in section \ref{subsecFixed} numerically.
Assuming that $\kappa_t$ will not deviate further than five standard deviations from its mean
$\mu_\kappa$, we find that one fixed point for the opinion index is indeed close to zero,
$s^* \sim \mathcal O (10^{-3})$, while the other two lie well outside of the domain of definition
$[-1, 1]$.
 
\subsection{Results and interpretation}
 
Figs. 1, 4 and 5 show the time dependence of the variables $P_t$, $s_t$, $\kappa_t$, $H_t$, 
$W_t$, $W^n_t$ and the time series of returns that are generated by numerical
solutions of the set (\ref{eqOpinionIndex}) to (\ref{eqMomentum}) for three different parameter values
for $\nu=1, 2$ and $0.5$ respectively, of the relative important of noise traders compared 
with rational investors in their price impact.

Fig.~1 corresponds to the situation where both groups have equal initial 
endowments ($\nu = 1$). One can observe a general positive log-price trend 
biasing upward a fluctuating random walk-like trajectory. The upward drift reflects 
a combination of the dividend gains, of the
rate of return paid by the risk-free asset as well as a component
resulting from the herding behavior of noise traders who 
tend intermittently to push prices in a kind of self-fulfilling prophecy or
convention \`a la Orl\'ean 
(Boyer and Orl\'ean 1992; Orl\'ean, 1994; Eymard-Duvernay et al. 2005).  

But the most striking aspect of the
price dynamics is the occurrence of four clearly identifiable bubbles
occurring within the chosen time interval, defined
by the transient explosive growth of the price $P_t$ followed by
sharp crashes bringing the prices back approximately to pre-bubble levels.
As seen from the second panel of Fig.~1 showing the opinion index dynamics of the noise traders, 
the bubbles are essentially driven by the noise 
traders. As described in section 4.3, the start of the growth of herding among noise
traders feeds the price dynamics, resulting in a larger price momentum (fourth panel), which 
amplifies herding, enhancing further the bubble growth and so on. 
One can observe in each bubble that the growth of the opinion index (or equivalently the
fraction of wealth invested in the risky asset) precedes and then 
accompanies the explosive price growth, as predicted by expression (\ref{eqExpP}).
The transient bubbles and their subsequent crashes are associated with
clustered volatility and the existence of outliers in the price momentum. 
During the bubbles, the wealth levels of noise traders and of
rational investors diverge. In the long run, noise traders outperform rational investors 
because they tend to invest more in the risky asset,
which exhibits higher average returns.

Fig.~2 presents a more detailed analysis of a typical bubble from the time series shown 
in Fig.~1, demonstrating the characteristic transient faster-than-exponential growth behavior
predicted theoretically in section 4.3.  For periods when $\kappa_t > p$, 
we may approximate the opinion index as exponentially growing:
\begin{equation}
s_t = s_1 (\alpha_1^t -1) ~,
\end{equation}
where $t$ runs over the growth period $[t_1, t_1 + \Delta T]$, with initial value 
$s_1 \equiv s_{t_1}$ and where $\alpha_1>1$ is an empirical effective multiplicative
factor, $\log \alpha_1$ being the effective growth rate of $s_t$.
 One can verify that the length $\Delta T$ of such a period is compatible with
our theoretical prediction (\ref{eqDeltaT}), which for our chosen parameters gives
$\Delta T = 20$. Bubbles with longer lifetimes are easily engineered 
in our framework by allowing $\kappa$ to remain close and higher than $p$
for longer times. Our model supports therefore the view that long-lived
bubbles may be associated with excess positive sentiments catalyzing
a herding propensity that is sustained and self-reinforcing (via the momentum
mechanism) over long periods.

Furthermore, the exponential growth in the opinion index results in
a faster-than-exponential growth of the price, as can be seen in the log-linear plot of
$P_t$. From expression (\ref{eqExpP}), we deduce
\begin{equation}
\log(P_t) = b_1 s_1 (\alpha_1^t - 1) + \log(P_0) ~,
\end{equation}
where $b_1 = b_{t_1}$, which fits well the transient
super-exponential price dynamics. These observations presented in Fig.~2 are in 
agreement with the theoretical derivation of section 4.3.
It is interesting to note also that the dynamics of $\kappa_t$, with its 
tendency to present a transient oscillatory behavior due to the interplay
between rare large excursions with the mean reversal of the
constrained random walk associated with the discrete Ornstein-Uhlenbeck
process, leads to an approximate  log-periodic behavior\footnote{Log-periodicity
here refers to transient oscillations with increasing local frequency. 
Formal mathematical definitions and illustrations can be found in (Sornette, 1998).} of the price
during its ascendency, which is similar to many observations
reported empirically (Sornette, 2003; Johansen and Sornette, 2010; 
Jiang et al., 2010; Yan et al., 2012; Sornette et al., 2013).

Fig.~3 presents three statistical properties of our generated price time series. Various
well-known stylized facts are matched by our model. First, we show the 
distribution of absolute values of the returns, which has a fat-tail 
$p(x)\sim x^{-1-\alpha}$ with exponent $\alpha = 3.0$, which is in the range
of accepted values in the empirical literature (de Vries, 1994; Pagan, 1996;
Guillaume et al., 1997; Gopikrishnan et al., 1998; Jondeau and Rockinger, 1999).
Furthermore, signed returns $R_t$ are characterized by a fast-decaying autocorrelation function,
which is consistent with an almost absence of arbitrage opportunities in the presence
of transaction costs. In contrast, the
absolute values  $|R_t|$ of returns have an autocorrelation function with longer memory
(Ding et al., 1993; Cont, 2007).

Figs.~4 and 5 present the same panels as in Fig.~1 but with $\nu = 2$ and $\nu = 1/2$, respectively.
Due to their larger relative weight compared to the case shown in Fig.~1, one can 
observe in Figs.~4 bubbles with stronger ``explosive'' trajectories. The wealth of noise traders fluctuates
widely, but amplifies to values that are many times larger than that of rational investors. This is due to the self-fulfilling
nature of the noise trader strategies that impact the price dynamics.
In contrast, Fig.~5 with $\nu = 1/2$ shows that the wealth of the rational investors remains high
for a long transient, even if in the long term the noise traders end up dominating the
price dynamics. The noise traders also transiently over-perform dramatically the 
rational investors during the bubbles. It is informative to observe that, even when they
are a minority ($\nu = 1/2$ shown in Fig.~5), the noise traders end up creating bubbles and crashes. Their 
influence progressively increases and their transient herding behavior becomes intermittently
destabilizing.

\subsection{Comparison with the dotcom bubble}

This section compares the insights obtained from the above theoretical 
and numerical analyses to empirical evidence on momenta and reversals 
in the period when the dotcom bubble developed.\footnote{The dotcom 
bubble (followed by its subsequent crash) is widely believed 
to be a speculative bubble, as documented by Ofek and Richardson (2003), 
Brunnermeier and Nagel (2004), and Battalio and Schultz (2006).}  
We study the characteristics of the share prices of Internet-related companies over the period 
from January 1, 1998 to December 31, 2002, which covers the period of the development
of the dotcom bubble and its collapse.
We use the list of 400 companies belonging to the Internet-related sector
that has been published by Morgan Stanley and has already been investigated by Ofek and Richardson (2003). 
The criteria for a company to be included in that list is that it must be considered a ``pure'' internet company,
i.e., whose commercial goals are associated exclusively to the Internet.
This implies that technology companies such as Cisco, Microsoft, and telecommunication firms, 
notwithstanding their extensive Internet-related businesses, are excluded. \par

Fig.~6 graphs the index of an equally weighted portfolio of the Internet stocks over the sample period of January 1998 to December 2002. The time evolution of the equally weighted portfolio of the Internet stocks is strikingly different
from that shown in figure 7 for the index of an equally weighted portfolio of non-Internet stocks over this same period. The two indexes are scaled to be 100 on January 2, 1998. The two figures illustrate clearly the widely held view that a divergence developed over this period between the relative pricing of Internet stocks and the broad market as a whole. In the two year period from early 1998 through February 2000, the internet related sector earned over 1300 percent returns on its public equity while the price index of the non-internet sectors rose by only 40 percent. However, these astronomical returns of the Internet stocks had completely evaporated by March 2001. 
Note how Fig.~6 is strikingly similar to the dynamics generated
by the theoretical model in the bubble regime shown at the end of the top panel of Fig.~5 ($\nu=0.5$).

\vskip 0.5cm
{\bf Table 1: Annual Returns for Internet and non-Internet stock indices}
\begin{center}
    \begin{tabular}{ r r r r r r }
    \hline
\multicolumn{6}{c}{}\\
    Year & 1998 & 1999 & 2000 & 2001  & 2002 \\ \hline
   Internet stock index  & $ 116.8 \%$ &  $ 815.6 \%$ &   $ - 875.9\%$ &  $ - 62 \%$ & $- 48.8 \%$  \\
              (per month)  & $ (9.7 \%) $ &  $ (68 \%)  $ &   $ (- 73 \%) $ &  $ (- 5.2 \%)$ & $ (- 4.1 \%)$  \\ \hline
   Non-internet stock index & $ 6.5\%$ & $17 \%$ &   $ - 9\%$ &  $ 3.6\%$ & $- 9 \%$  \\
                     (per month) & $ (0.5\%)$ & $(1.4 \%)$ &   $ (- 0.8\%)$ &  $ (0.3 \%) $ & $(- 0.7 \%)$  \\ \hline
     \end{tabular}
\end{center}
\vskip 0.5cm

We now focus our attention on the profitability of the momentum strategies studied by Jegadeesh and Titman (1993, 2001) and others. Table 1 provides some descriptive statistics about annual returns of 
the Internet-stock index versus of the non-Internet stock index from the beginning of 1998 to the end of 2002. In the 12 months of 1998, the annual cumulative return of the Internet stock index was 117 percent, while that of the non-Internet stock index was 6.5 percent. In the 12 months of 1999, the  annual cumulative return of the Internet stock index surged to 816 percent, and that of the non-Internet stock index increased to 16.6 percent. The Internet stock index clearly outperformed the non-Internet stock index by 800  percent in 1999. This implies a strong profitability of momentum strategies applied to the Internet stocks over the period of the dotcom bubble. However, after its burst in March 2000, the return of the Internet stocks sharply declined, from 2000 to 2002. In the 12 months of 2000, the annual return of the internet-stock index fell to - 876 percent, followed
by  - 62 percent and - 49 percent in 2001 and in 2002, respectively. On the other hand, the annual returns of the non-Internet stock index in the period from 2000 to 2002 remain modest in amplitude at - 9 percent, 3.6 percent and - 9 percent, respectively. After the bust of the dotcom bubble, the Internet stocks 
continued to underperform the non-Internet stocks. 

Table 2 shows the cumulative returns for the Internet stock index and for the non-Internet stock index in the five years from the beginning of 1998 to the end of 2002. The cumulative return of the Internet stock index in the first 24 months of the holding period is 932.5 percent, but the cumulative returns ends at
the net loss of  - 54.2 percent over the five year holding period. In contrast, the cumulative returns of the non-Internet stock index over the same five year holding period is 8.6 percent. 

These figures can be reproduced by our simulations, and are visualized by the extremely good
performance of our noise traders during the bubble phases, as shown in the fifth panels (from the top) of
Figs. 1, 4 and 5.

\bigskip
\bigskip

{\bf Table 2: Cumulative Returns for Internet and non-Internet stock indices}
\begin{center}
    \begin{tabular}{ r r r r r r }
    \hline
\multicolumn{6}{c}{}\\
    Year & 1998 & 1999 & 2000 & 2001  & 2002 \\ \hline
   Internet stock index  & $ 116.8 \%$   &  $ 932.5 \%$   &   $ 56.6 \%$   &  $ - 5.4 \%$   & $- 54.2 \%$  \\
              (per month)  & $ (9.7 \%)$ &  $ (38.9 \%)$ &   $ (1.6 \%)$ &  $ (- 0.1 \%)$ & $ (- 0.9 \%)$  \\ \hline
   Non-internet stock index & $ 6.5\%$ & $23.1 \%$ &   $ - 14 \%$ &  $ 17.6 \%$ & $ 8.6 \%$  \\
                     (per month) & $ (0.5 \%)$ &  $ (1.0 \%)$ &   $ (0.4 \%)$ &  $ (0.4 \%)$ & $ (0.1 \%)$  \\ \hline
     \end{tabular}
\end{center}

In summary, these empirical facts constitute strong evidence
for the Internet stock for momentum profit at 
intermediate time scales of about two years and reversals at longer time scales of about 5 years.
These empirical facts confirm for this specific bubble and crash period
the general evidence documented by many researchers (e.g. Jegadeesh and Titman, 1993, 2001).
They are consistent with the stylized facts described by the model that predict that the momentum profits will eventually reverse in cycle bubbles and crashes as illustrated above. The quantitative comparison
between the empirical data and our simulations suggest that noise traders do not need to be
a majority, as their superior performance during the bubble make them dominate eventually
utterly the investment ecology.

 \section{Conclusions}
 
We have introduced a model of financial bubbles 
with two assets (risky and risk-free),
in which rational investors and noise traders co-exist. Rational investors
form expectations on the return and risk of a risky asset and 
maximize their constant relative risk aversion expected utility with respect to their portfolio allocation. 
Noise traders are subjected to social
imitation and follow momentum trading. 

In contrast to various previous models, agents do not switch between investment strategies.
By keeping track of their wealth levels, we still observe the formation of endogenous
bubbles and match several stylized facts of financial markets such as a fat-tail
distribution of returns and volatility clustering. In particular, we observe transient faster-than-exponential 
bubble growth with approximate log-periodic behavior. Although faster-than-exponential growth at the 
beginning of a bubble has been found 
for many bubbles in real markets and recent lab experiments, it has been hardly discussed
in the literature. Our model is one of the first offering a transparent analytical explanation for
this stylized fact.

To the important question of whether and when rational 
investors are able to stabilize financial markets by arbitraging 
noise traders, our analysis suggests that noise traders may 
eventually always lead to the creation of bubbles, given sufficient time,
if a mechanism exists or some sentiment develops that increase
their propensity for herding.

The model has been found to account well for the behavior of traders and for the price
dynamics that developed during the dotcom bubble in 1995-2000.
Momentum strategies have been shown to be transiently profitable, supporting
the hypothesis that these strategies enhance herding behavior.

\vskip 0.5cm  
{\bf Acknowledgements}: We are grateful to Yannick Malevergne
for stimulating remarks on an earlier version of the manuscript.

\clearpage
 
\begin{quote}
\centerline{\includegraphics[width=1.1\textwidth]{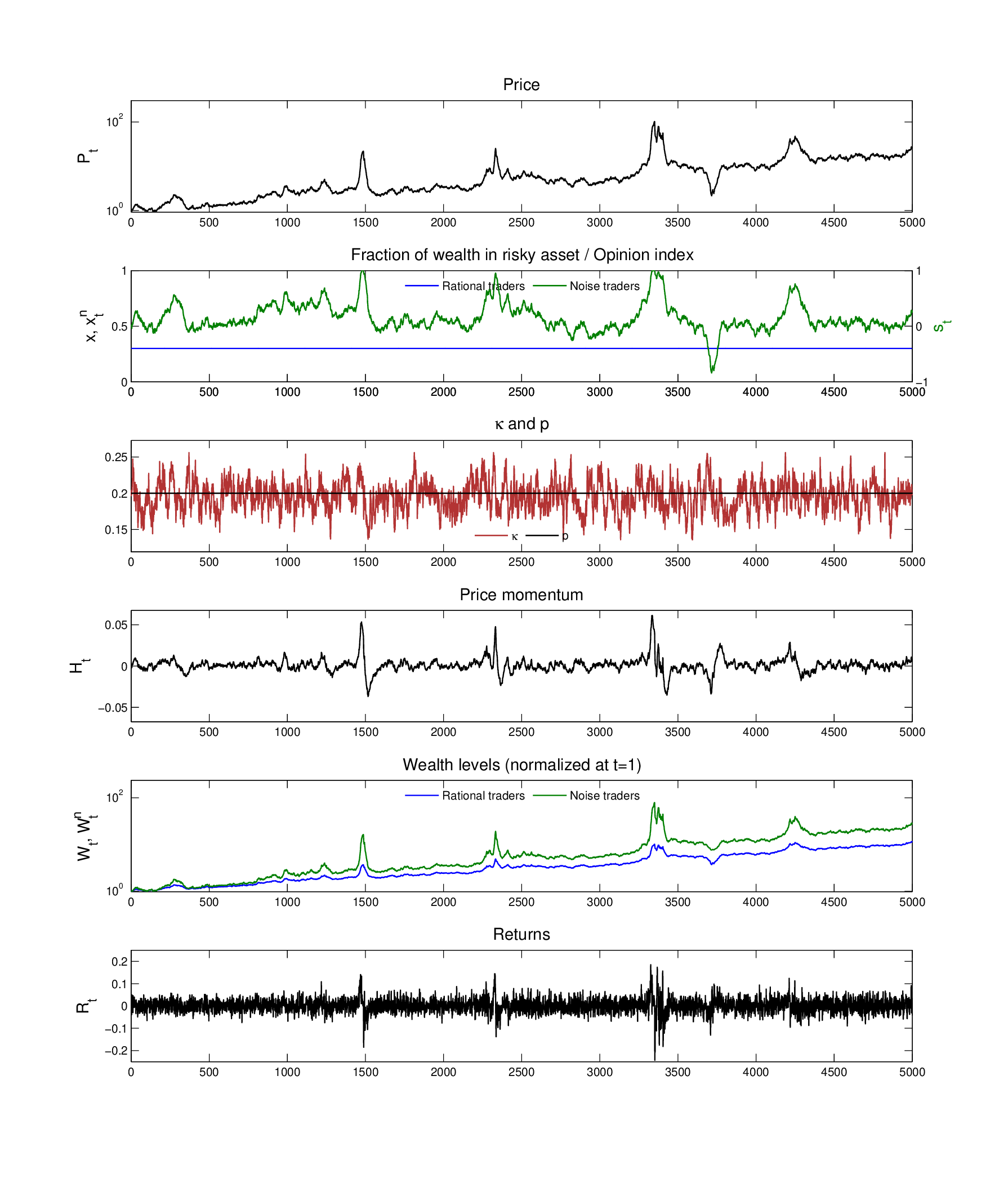}}
\vskip -1cm
{\bf Fig.~1:} \small{Time dependence of the variables $P_t$ (log-scale), $s_t$, $\kappa_t$, $H_t$, 
$W_t$, $W^n_t$ (both in log-scale) and the time series of returns that are generated by numerical
solutions of the set (\ref{eqOpinionIndex}) to (\ref{eqMomentum}) for 
the value $\nu=1$ of the relative important of noise traders compared 
with rational investors in their price impact at the origin of time. }
\end{quote}

\begin{quote}
\centerline{\includegraphics[width=1.0\textwidth]{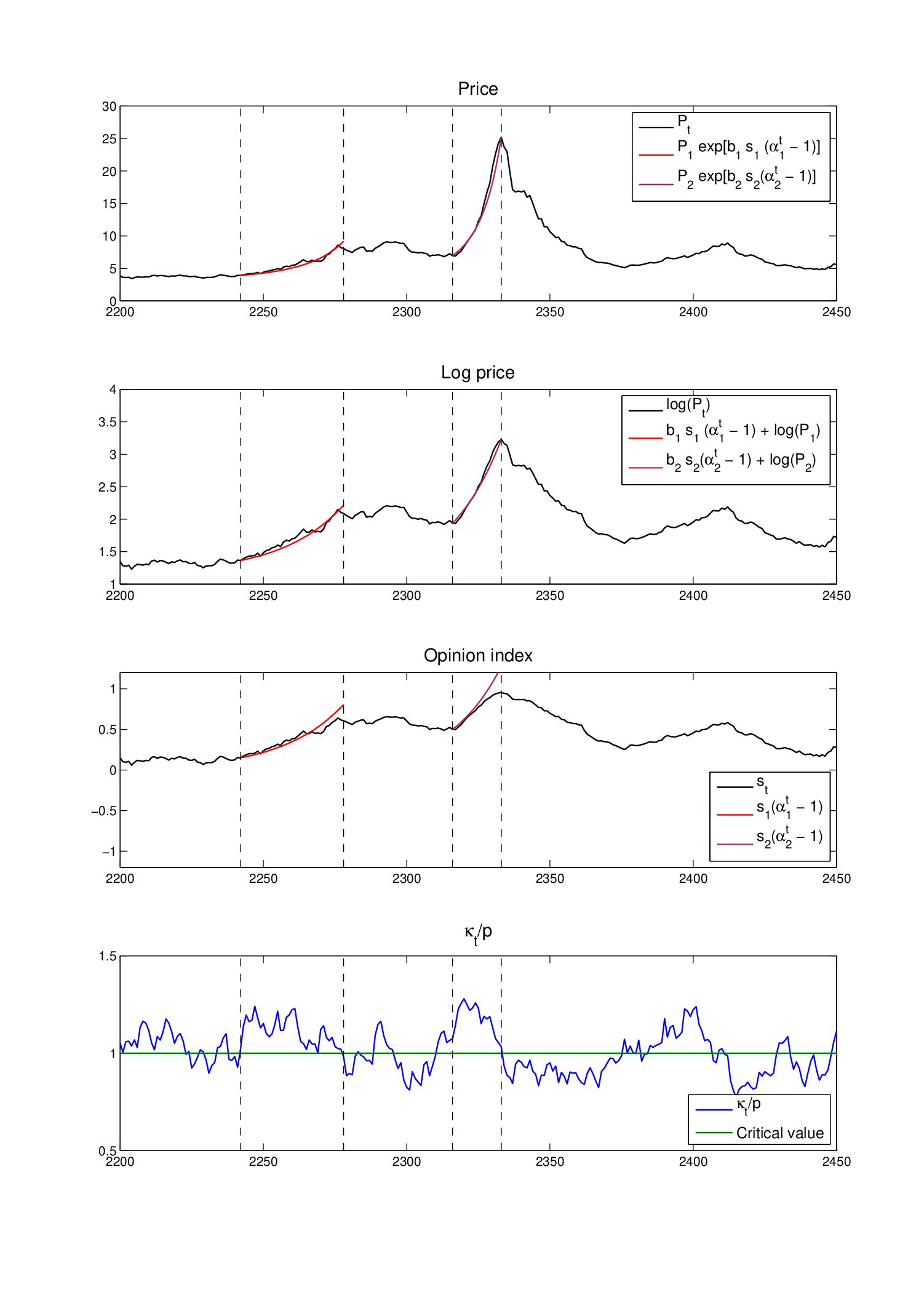}}
\vskip -1cm
{\bf Fig.~2:} \small{Zoom of Fig.~1 for the price $P_t$, log-price $\ln P_t$, opinion index $s_t$
and imitation parameter $\kappa_t$ in units of the random component strength $p$
as a function of time around bubbles. The panels show the behavior of these variables
for typical bubbles, demonstrating the characteristic transient faster-than-exponential growth behavior.}
\end{quote}

\begin{quote}
\centerline{\includegraphics[width=0.67\textwidth]{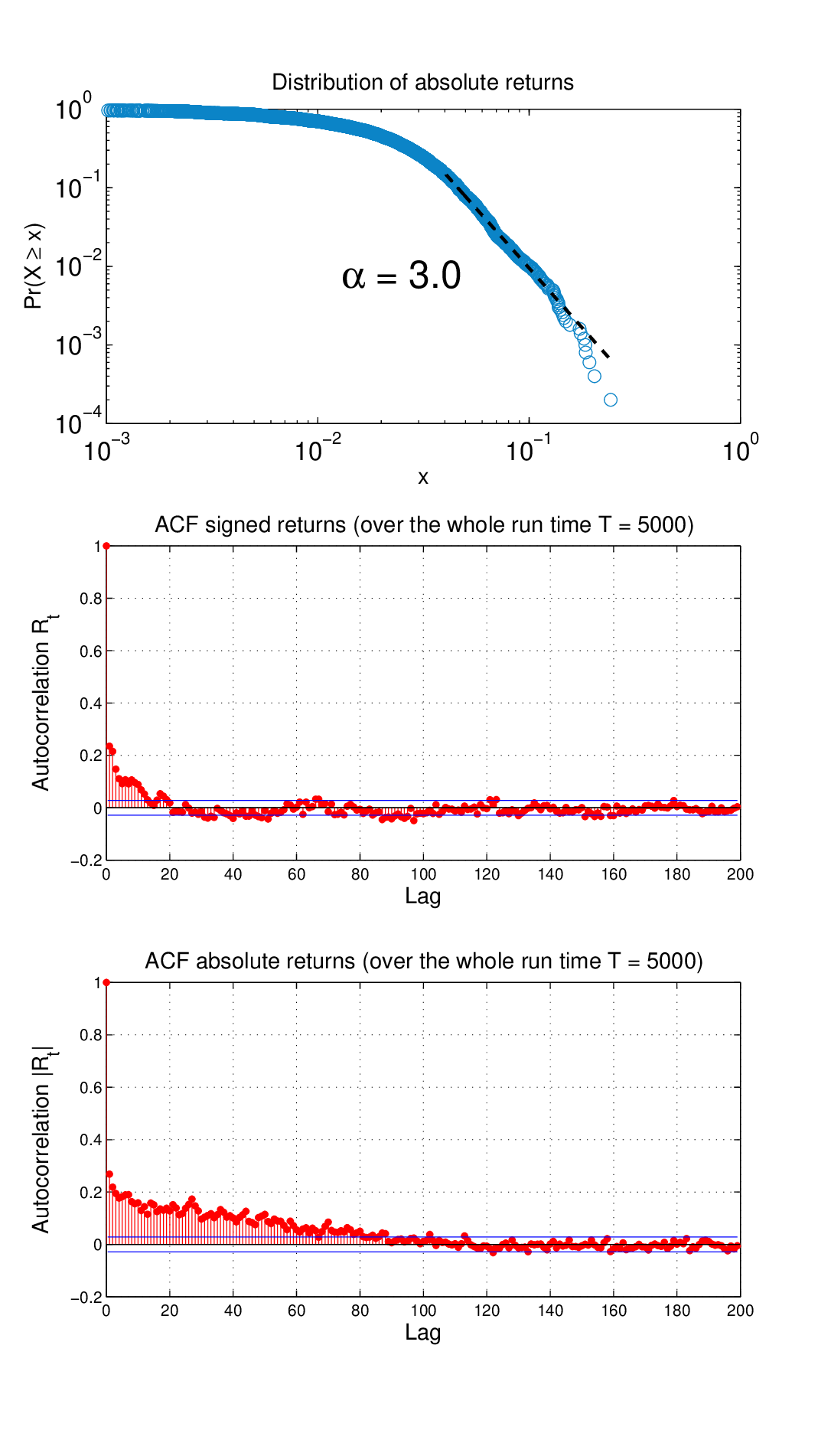}}
\vskip -1.5cm
{\bf Fig.~3:} \small{Top panel: complementary cumulative
distribution function of absolute values of the returns in log-log, where
the straight dashed line qualifies a fat-tail 
$p(x)\sim x^{-1-\alpha}$ with exponent $\alpha = 3.0$;
Middle panel: auto-correlation function of the signed returns $R_t$;
Lower panel: autocorrelation function of the absolute values  $|R_t|$ of returns.
The parameters are the same as in Fig.~1.
}
\end{quote}

\begin{quote}
\centerline{\includegraphics[width=1.1\textwidth]{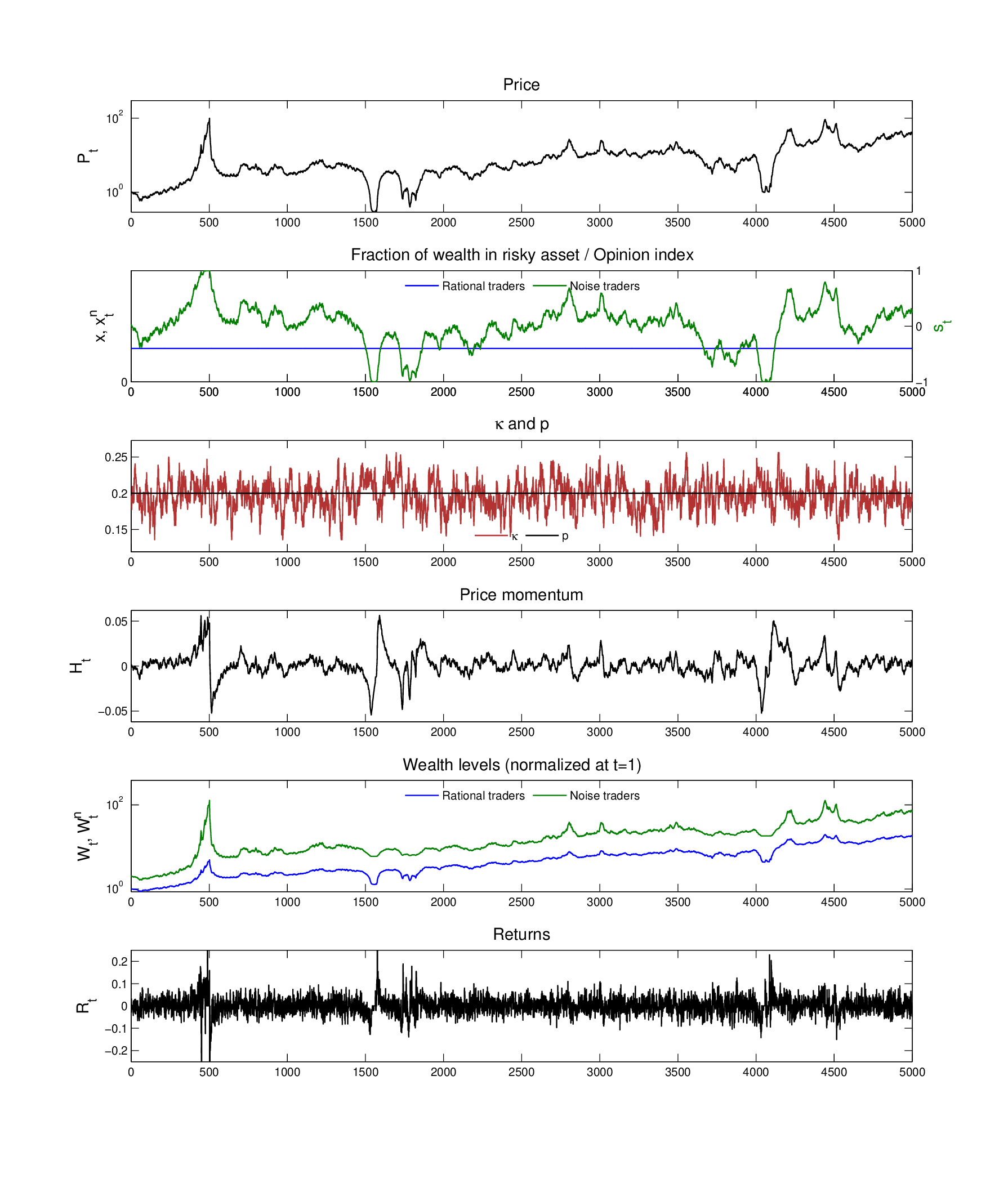}}
{\bf Fig.~4:} \small{Same as Fig.~1 for $\nu=2$.}
\end{quote}

\begin{quote}
\centerline{\includegraphics[width=1.1\textwidth]{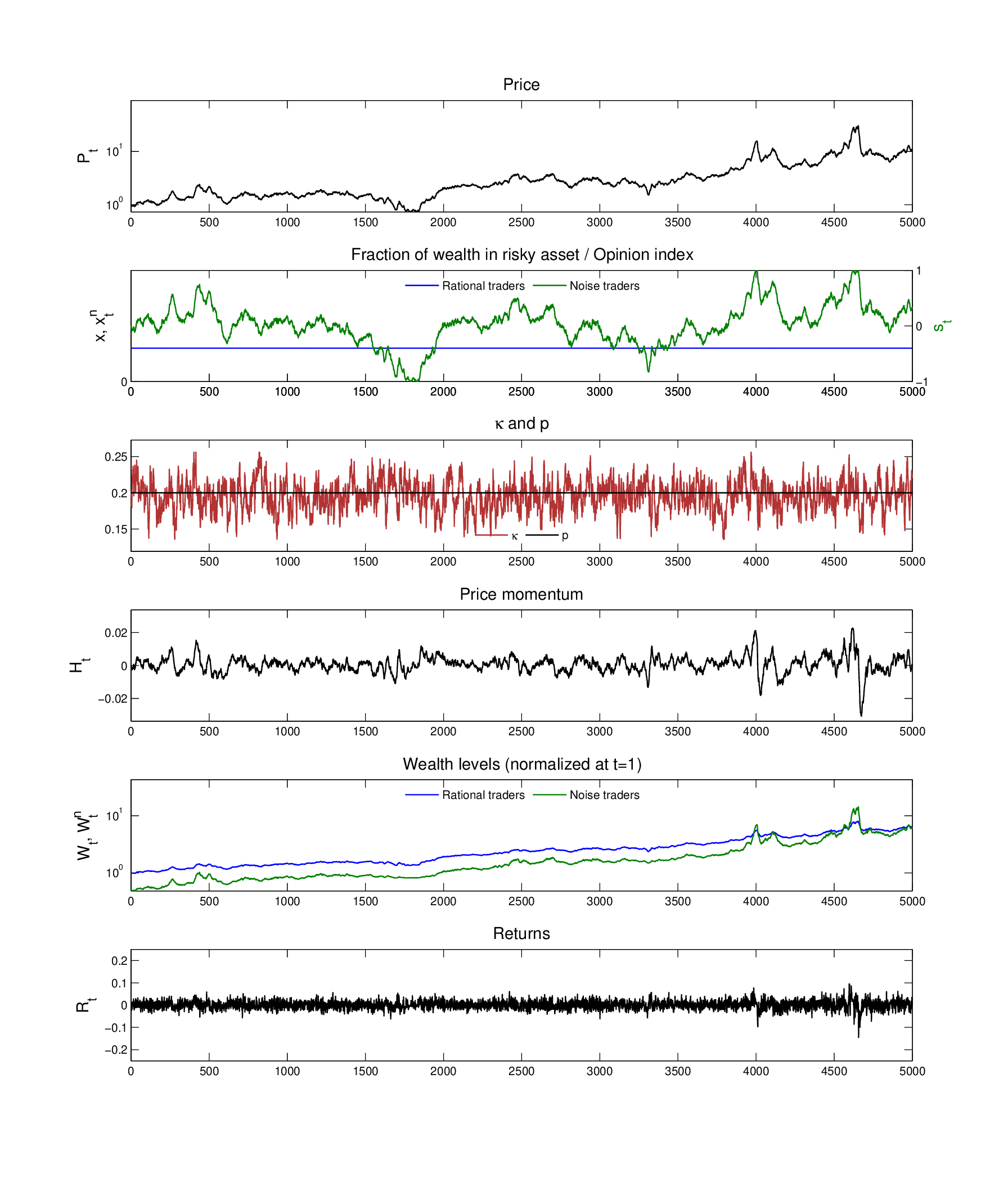}}
{\bf Fig.~5:} \small{Same as Fig.~1 for $\nu=0.5$.}
\end{quote}

\begin{quote}
\centerline{\includegraphics[width=0.75\textwidth]{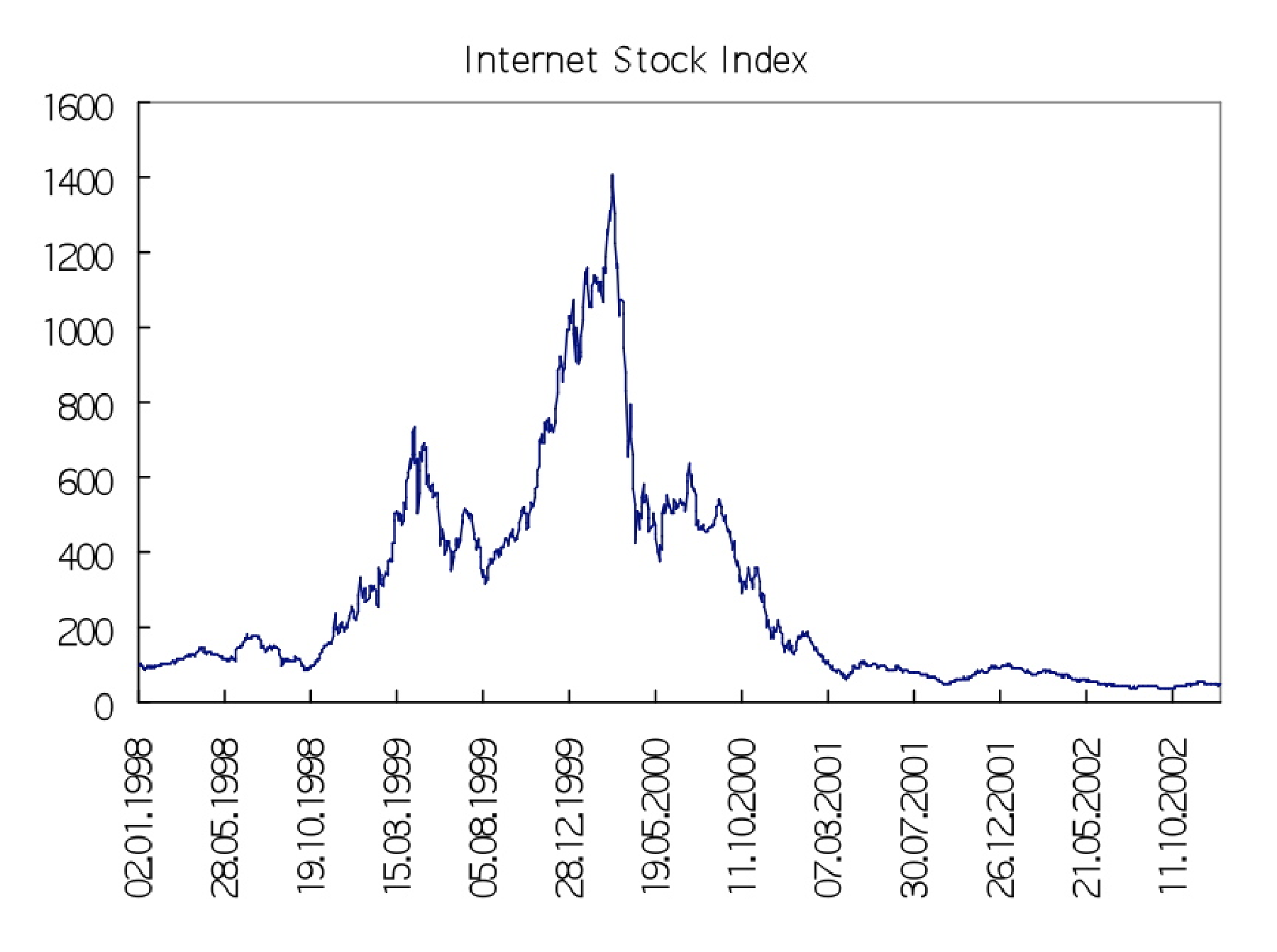}}
{\bf Fig.~6:} \small{The equally weighted Internet stock index  for the period 1/2/1998-12/31/2002. The index is scaled to be 100 on 1/2/1998. }
\end{quote}

\begin{quote}
\centerline{\includegraphics[width=0.75\textwidth]{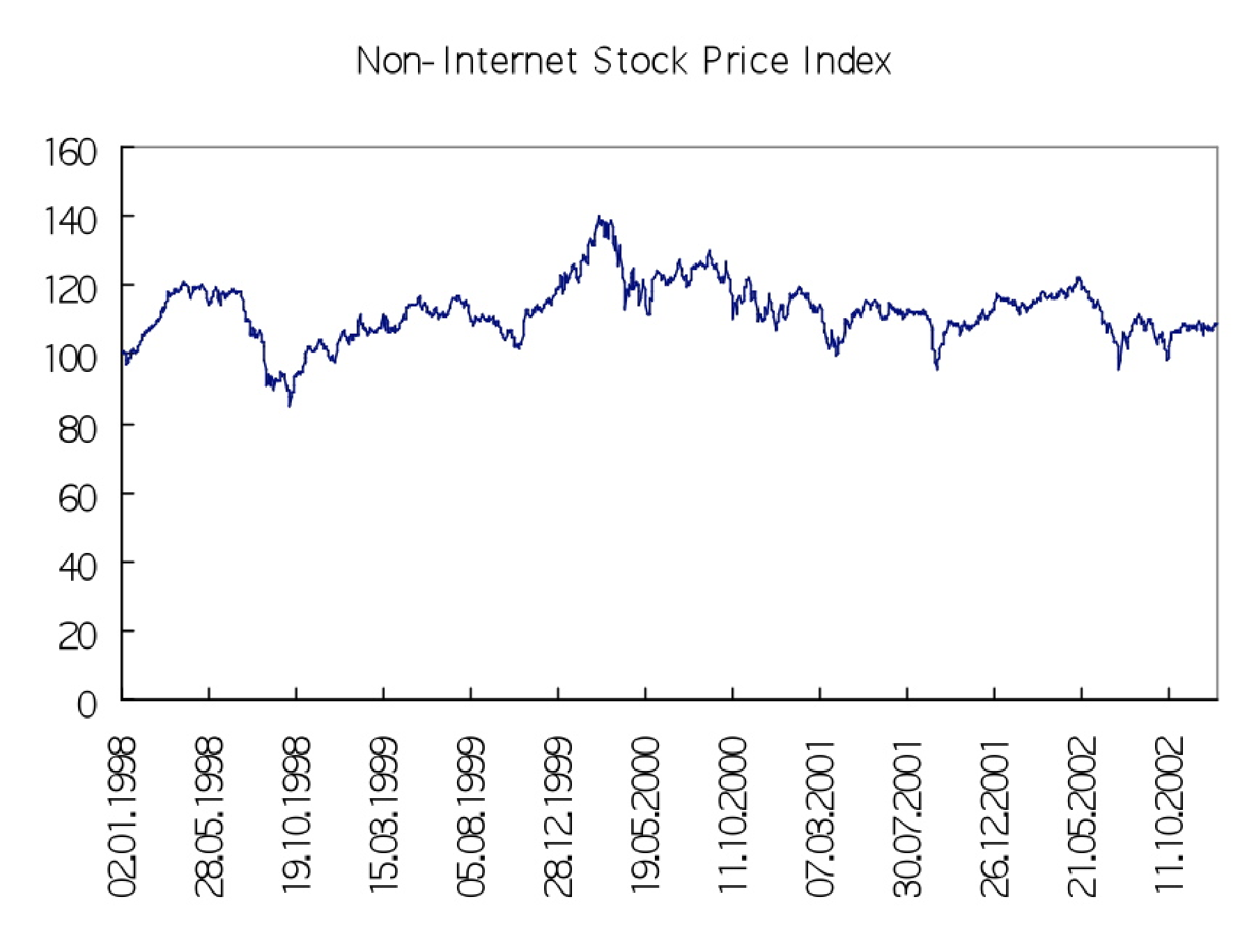}}
{\bf Fig.~7:} \small{ The equally weighted non-Internet stock index for the period 1/2/1998-12/31/2002. The index is scaled to be 100 on 1/2/1998. }
\end{quote}

\end{document}